\documentclass[a4paper,onecolumn,11pt,unpublished]{quantumarticle}
\pdfoutput=1
\usepackage[T1]{fontenc}
\usepackage[utf8]{inputenc}
\usepackage{amsmath,amssymb,bm,microtype}
\usepackage{booktabs,array,tabularx}
\usepackage[numbers,sort&compress]{natbib}
\usepackage{xcolor,url,graphicx}
\usepackage{subcaption}
\usepackage[normalem]{ulem} 

\usepackage[hidelinks]{hyperref}
\newcommand{\R}{\mathbb R}

\newcommand{\ket}[1]{\lvert#1\rangle}
\newcommand{\bra}[1]{\langle#1\rvert}
\newcommand{\JS}{D_{\mathrm{JS}}}

\usepackage{etoolbox}
\makeatletter
\pretocmd{\@maketitle}
  {\centering\toggletrue{@titlepage}}
  {}{}
\makeatother
\title{Conditional Quantum Priors for Chaotic Fluid Dynamics}

\author{\parbox{\textwidth}{\centering%
      \begin{tabular}{@{}c@{\hspace{1.5em}}c@{\hspace{1.5em}}c@{}}
      Shao-Bo Zhou$^{1}$ &
      Yu-Ze Zhu$^{1}$ &
      Hang Song$^{2}$ \\[0.3em]
      Xiang Zhao$^{1,\dagger}$ &
      Lin Yang$^{1,\dagger}$ &
      Xian-Min Jin$^{1,2,3,4,5,\dagger}$
    \end{tabular}
    \\[0.6em]
  $^{1}$TuringQ Co., Ltd., Shanghai 200240, China
  \\[0.3em]
  $^{2}$Center for Integrated Quantum Information Technologies (IQIT), \\
  School of Physics and Astronomy and State Key Laboratory of \\
  Photonics and Communications, \\
  Shanghai Jiao Tong University, Shanghai 200240, China
  \\[0.3em]
  $^{3}$Chip Hub for Integrated Photonics Xplore (CHIPX), \\
  Shanghai Jiao Tong University, Wuxi 214000, China
  \\[0.3em]
  $^{4}$Atomology Co., Ltd., Shanghai 200240, China
  \\[0.3em]
  $^{5}$Hefei National Laboratory, Hefei 230088, China
}}

\begin{document}
\maketitle
\begingroup
\renewcommand{\thefootnote}{\fnsymbol{footnote}}
\footnotetext[2]{Corresponding authors:
\texttt{zhaoxiang@turingq.com},
\texttt{yanglin@turingq.com},
\texttt{xianmin.jin@sjtu.edu.cn}.}
\endgroup

\begin{abstract}
Neural networks that predict chaotic dynamics can reproduce short trajectories while misrepresenting statistics over long times.
Quantum priors represent selected dynamical statistics through quantum generative models and provide statistical constraints for classical neural networks.
Training a separate prior for each value of a system parameter,
such as the Reynolds number, does not explicitly model how these
statistics vary with that parameter.
We learn a conditional quantum prior that approximates selected statistics of an invariant measure at unseen parameter values within the training range, without retraining.
For Lorenz-63, the prior reduces mean Jensen--Shannon divergence by 56.9\% relative to a classical multilayer perceptron with a comparable number of weights across 96 unseen parameter values.
Across 101 values of the same parameter, the shared prior uses
84.7\% fewer trainable weights than would be required by a separate
quantum neural network at each value. We then use the learned prior to regularize classical neural networks, encouraging their autoregressive rollouts to better reproduce long-term statistics. For Kolmogorov flow at Reynolds number $Re=1000$, prior regularization reduces forecast mean squared error by 66.0\% over 318 steps relative
to the classical baseline without prior regularization.
We also implement a photonic quantum prior for noisy intermediate-scale quantum (NISQ) devices, achieving a mean Jensen--Shannon divergence of $4.58\times10^{-3}$ for Lorenz-63 and $2.59\times10^{-3}$ for Kolmogorov flow at unseen parameter values.
\end{abstract}

\section{Introduction}\label{sec:introduction}


Chaotic fluid dynamics are sensitive to initial conditions~\citep{lorenz1963,lucas2015}.
Differential equations describe how these systems evolve.
Ordinary differential equations (ODEs) describe how a finite set of state variables changes in time, as in the Lorenz model of convection~\citep{lorenz1963}.
Partial differential equations (PDEs) additionally describe spatial variation, as in the Kuramoto--Sivashinsky equation and the Navier--Stokes equations that govern Kolmogorov flow~\citep{kuramoto1976,lucas2015}.
In chaotic regimes, small prediction errors can grow rapidly, making it difficult for neural networks to follow individual trajectories over long periods~\citep{lorenz1963,li2022chaotic}.
Reproducing the statistics of the flow therefore provides a complementary measure of prediction quality when individual trajectories diverge~\citep{young2002,jiang2023invariant}.


Neural surrogate models learn system dynamics from simulations or measurements~\citep{li2021fno,lusch2018}.
Examples include DeepONet and Fourier neural operators, which learn mappings between input and solution functions~\citep{lu2021deeponet,li2021fno}.
We use \emph{predictor} to denote a classical neural network that advances the system state.
Repeated prediction gives an autoregressive rollout, with each predicted state serving as the next input.
However, fitting short trajectories alone does not ensure that the model reproduces the statistics over long times~\citep{li2022chaotic,jiang2023invariant}.
Physics-informed machine learning supplements data fitting with physical constraints~\citep{karniadakis2021}.
For chaotic systems, statistical regularization encourages predictions to reproduce invariant statistics, such as the mean energy at different spatial scales in a stationary flow~\citep{young2002,jiang2023invariant}.
Jiang et~al. penalize differences in invariant statistics~\citep{jiang2023invariant}, while Dynamics Stable Learning by Invariant Measure (DySLIM) compares distributions of observed and predicted states~\citep{schiff2024}.
Quantum-informed machine learning (QIML) takes a related approach by first learning a statistical distribution with a separate quantum generative model~\citep{wang2026}.
This learned distribution serves as a fixed \emph{prior} during classical predictor training.
The resulting hybrid quantum--classical model uses quantum measurements to represent statistics and a classical neural network to predict trajectories~\citep{biamonte2017,cerezo2021,wang2026}.


Dynamical statistics depend on system parameters such as the Reynolds
number.
A prior learned at a single parameter value, as in
QIML~\citep{wang2026}, may not accurately represent the statistics
at another.
Extending this approach to multiple parameter values requires a
separate optimization and set of trainable weights at each value.
The resulting independent priors do not learn a shared dependence
of the statistics on the system parameter.


To address this limitation, we develop a \emph{conditional quantum
prior} that learns how selected dynamical statistics depend on a system parameter. We investigate two implementations: a graph quantum circuit Born machine (\emph{Graph QCBM}) and a photonic quantum neural network (\emph{Photonic QNN}). The Graph QCBM combines parameter encoding with quantum evolution on a graph, while the Photonic QNN uses measured optical responses with trainable classical components. Both implementations use a shared set of trainable weights across parameter values. Once trained, each model provides statistical references at parameter values not used in prior training, without retraining. These references can then regularize classical predictors at the corresponding parameter values.

Superconducting qubits and trapped ions provide experimental platforms for Born machines~\citep{hamilton2019,benedetti2019}.
However, gate errors limit the circuit depth that noisy intermediate-scale quantum (NISQ) devices can execute reliably~\citep{preskill2018,cerezo2021}.
We therefore investigate the Photonic QNN as an alternative for experiments on NISQ hardware.
Photonic circuits mix optical modes with beam splitters and phase shifters, producing detection probabilities that vary with the encoded input~\citep{gan2022}.
The Photonic QNN combines responses measured on this hardware with trainable classical readouts and interactions to learn conditional probability distributions.

Our main contributions are:
\begin{enumerate}
    \item \textbf{Parameter efficiency.} A single conditional quantum prior shares trainable weights across different values of the system parameter, avoiding separate fits. For Lorenz-63, a prior trained at five values of $\rho$ provides statistical references across 101 parameter values, using 84.7\% fewer trainable weights than would be required by fitting a separate QIML quantum neural network at each value.
    \item \textbf{Prior accuracy.} The Graph QCBM achieves mean Jensen--Shannon divergence $0.5336\times10^{-3}$ across 96 unseen Lorenz-63 parameter values. This is 56.9\% lower than a classical multilayer perceptron with a comparable number of weights.
    \item \textbf{Improved forecasts over long horizons.} Regularization with the fixed prior reduces mean squared error by 66.0\% over 318 steps for Kolmogorov flow at $Re=1000$, relative to the same predictor trained without the prior.
\end{enumerate}

\section{Methods}\label{sec:methods}

\begin{figure}[h!]
\centering
\begin{subfigure}{\linewidth}
\centering
\includegraphics[width=\linewidth]{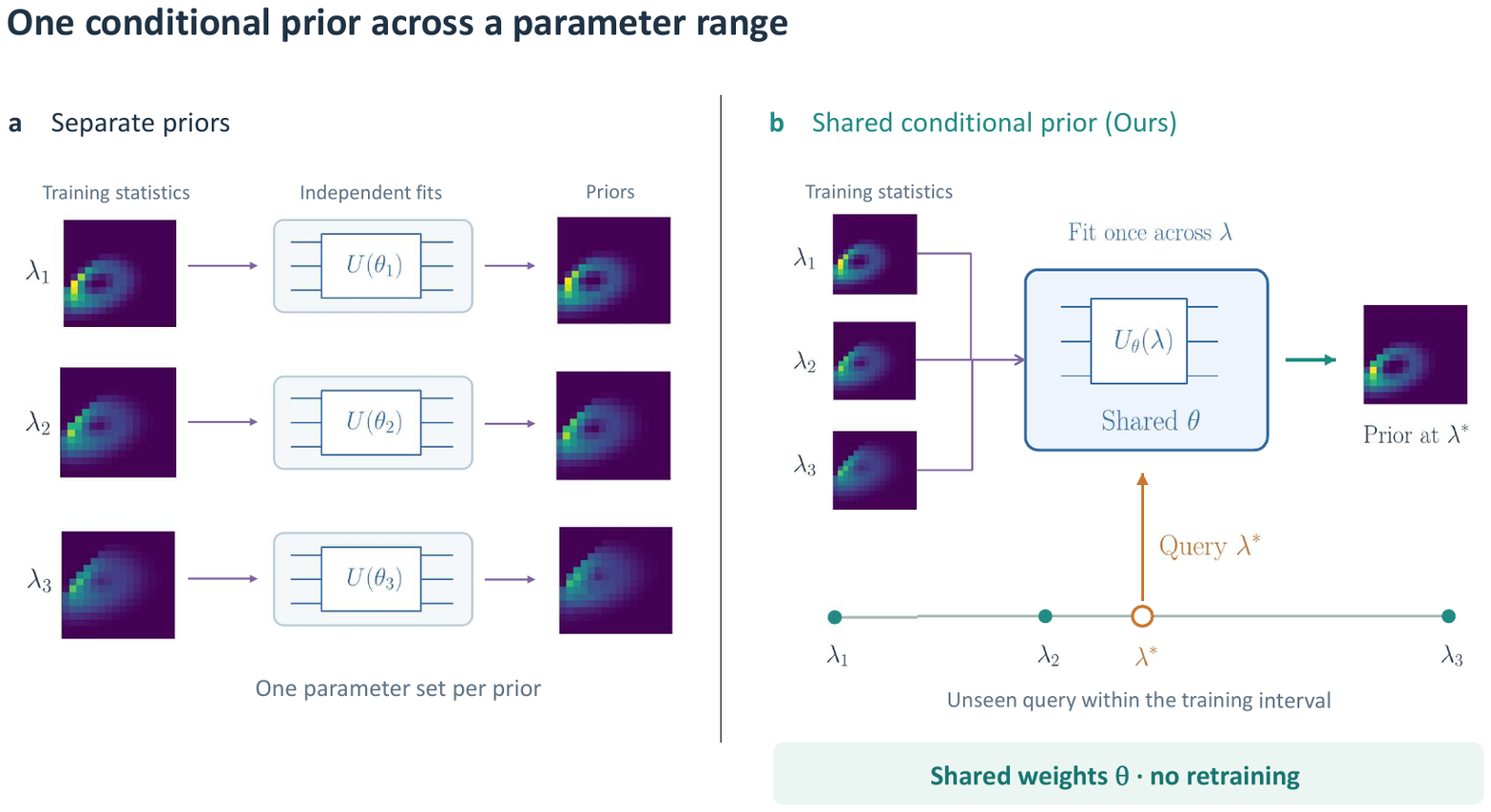}
\phantomcaption
\label{fig:prior-sharing}
\end{subfigure}
\par\medskip\vspace{-25mm}
\begin{subfigure}{\linewidth}
\centering
\includegraphics[width=\linewidth]{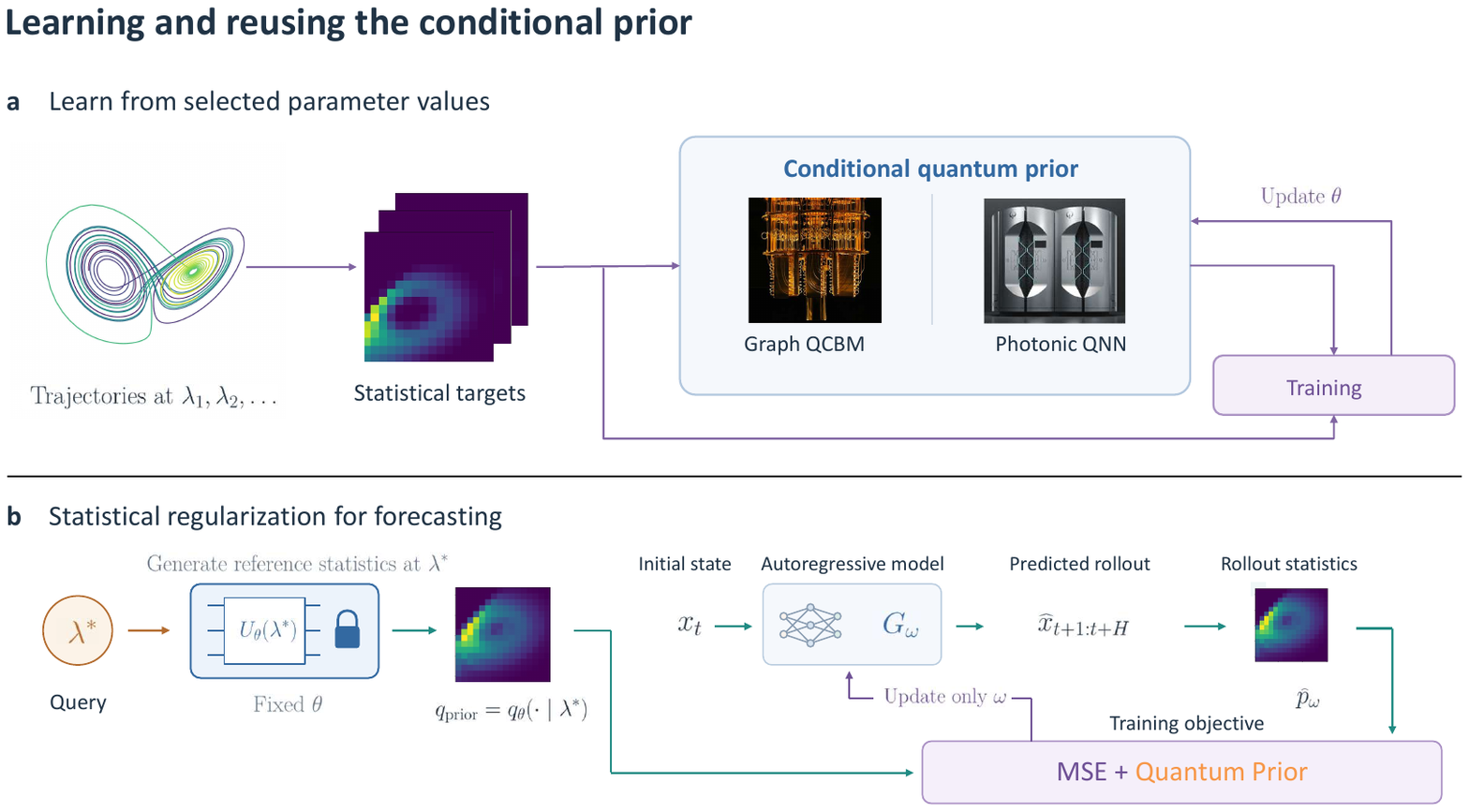}
\phantomcaption
\label{fig:prior-workflow}
\end{subfigure}
\vspace{-15mm}
\caption{Overview of the conditional quantum prior framework.
\textbf{A (top).} A comparison of independent priors and our shared conditional prior.
\textbf{B (bottom).} Learning a conditional prior
from trajectory statistics and using it to regularize
a classical predictor.}
\label{fig:prior-overview}
\end{figure}

Our framework uses a learned statistical prior to
guide the training of a classical predictor.
The prior is trained on statistical distributions
estimated from trajectories at several system
parameter values.
Unlike independently trained priors, the conditional
prior can be reused at different conditions without
retraining.
At a chosen parameter value, it provides reference
statistics against which the statistics of predicted
trajectories are compared.
The predictor is also trained to reproduce observed
state transitions, while the prior remains fixed.
Figure~\ref{fig:prior-overview} provides an overview
of the shared conditional prior and the
training workflow.

\subsection{Invariant Measures and Statistical Targets}\label{sec:targets}
We begin by defining the statistical targets
obtained from the underlying dynamics.
Consider sampled dynamics at a fixed system parameter $\lambda$,
\begin{equation}
 x_{t+1}=F_\lambda(x_t),
 \label{eq:dynamics}
\end{equation}
where $x_t\in\mathcal X$ is the complete system state
and $\lambda\in\mathbb R$ denotes a scalar system parameter.
A condition fixes the value of $\lambda$ in the governing equations.
At a fixed condition, the classical predictor $G_\omega$ approximates $F_\lambda$ with weights $\omega$ learned from trajectory data~\citep{lusch2018,wang2026}.
Training minimizes a data loss that measures agreement between predicted and observed states.
Starting from an observed state $x_t$, repeated applications generate an autoregressive rollout:
\begin{equation}
 \widehat x_t=x_t,
 \qquad
 \widehat x_{t+h}=G_\omega(\widehat x_{t+h-1}),
 \qquad h=1,\ldots,H,
 \label{eq:rollout}
\end{equation}
where $H$ is the forecast horizon and $\widehat x_{t+h}$ denotes a predicted state.

Although the predictor is trained to reproduce observed state transitions, accurate short-term predictions do not necessarily guarantee the preservation of long-term statistics in autoregressive rollouts. To characterize these statistics, we consider an invariant probability measure of the underlying dynamics. A probability measure $\mu_\lambda$ is invariant if $x\sim\mu_\lambda$ implies $F_\lambda(x)\sim\mu_\lambda$~\citep{young2002}.
We consider the invariant measure \(\mu_\lambda\) associated with the statistically stationary regime explored by the observed trajectories.

For an integrable observable $\mathcal O:\mathcal X\to\R$, we write its expectation under $\mu_\lambda$ as the invariant statistic:
\begin{equation}
 s_\lambda[\mathcal O]
 =\int_{\mathcal X}\mathcal O(x)\,\mu_\lambda(\mathrm dx).
 \label{eq:invariant-statistics}
\end{equation}
Since $F_\lambda$ preserves $\mu_\lambda$, this expectation satisfies $s_\lambda[\mathcal O\circ F_\lambda]=s_\lambda[\mathcal O]$, although individual values $\mathcal O(x_t)$ can fluctuate along a trajectory.
Under ergodicity, time averages converge to these expectations for $\mu_\lambda$-almost every initial state~\citep{young2002}.




To obtain finite-dimensional statistical targets for conditional prior learning, we select a collection of \(M\) nonnegative measurable functions \(\mathcal O_i:\mathcal X\to[0,\infty)\), indexed by \(i=1,\ldots,M\). These functions encode the statistical quantities of interest and are constructed according to the target type. For occupancy distributions, \(\mathcal O_i\) is an indicator function associated with a prescribed bin, and its invariant expectation gives the probability of that bin. For spectral targets, \(\mathcal O_i\) measures the power associated with a prescribed group of Fourier modes, and its invariant expectation gives the corresponding spectral weight. The specific constructions for each dynamical system are given in Section 3.1.

For each condition \(\lambda\), we define the invariant statistical weights as

\begin{equation}
w_\lambda(i)=s_\lambda[\mathcal O_i]
=\int_{\mathcal X}\mathcal O_i(x)\,\mu_\lambda(dx).
\end{equation}

Assuming that their total weight is positive, we define the normalized statistical target

\begin{equation}
p_\lambda(i)=
\frac{w_\lambda(i)}
{\sum_{j=1}^{M}w_\lambda(j)},
\qquad i=1,\ldots,M.
\end{equation}

The resulting vector \(\mathbf p_\lambda\) defines a discrete probability distribution over the \(M\) indices. For occupancy distributions, its entries represent bin probabilities when the bins form a complete partition. For spectral distributions, they represent the relative contributions of different Fourier mode groups to the total included spectral power. In the latter case, normalization discards information about the total spectral power.

Given an observed trajectory \(\{x_t\}_{t=1}^{T}\), we estimate the statistical weights by

\begin{equation}
\hat w_\lambda(i)=
\frac{1}{T}\sum_{t=1}^{T}\mathcal O_i(x_t),
\end{equation}
and obtain the empirical target \(\hat{\mathbf p}_\lambda\) by normalizing these estimates. Under the corresponding ergodicity and integrability assumptions, these empirical estimates converge to their invariant counterparts. For each system, we use fixed bin definitions or Fourier mode groupings across all parameter conditions to ensure comparability of the targets.

\subsection{Conditional Prior Learning}
A conditional quantum prior represents the dependence of the statistical
target on the system parameter through a single shared set of trainable
weights $\theta$:
\begin{equation}
 \lambda\longmapsto q_\theta(\cdot\mid\lambda),
 \qquad q_\theta(i\mid\lambda)\geq0,
 \qquad \sum_{i=1}^M q_\theta(i\mid\lambda)=1,
 \label{eq:interface}
\end{equation}
where $\lambda$ is the conditioning input and $i$ indexes the prescribed
categories or bins.

We use the Jensen--Shannon (JS) divergence~\citep{lin1991} to train the
Graph QCBM and Photonic QNN, evaluate their learned distributions, and
regularize classical predictors.
For probability distributions $p$ and $q$, it is defined as
\begin{equation}
 \JS(p,q)=
 \frac12\sum_i p_i\log\frac{p_i}{m_i}
 +\frac12\sum_i q_i\log\frac{q_i}{m_i},
 \qquad m_i=\frac{p_i+q_i}{2}.
 \label{eq:js}
\end{equation}
We use natural logarithms and assign zero contribution to terms with
zero numerator.

Let $\Lambda_{\rm train}$ denote the set of system parameter values
used to fit the prior.
We call these values \emph{seen} conditions and all other admissible
values \emph{unseen} conditions, with both terms referring specifically
to prior training.
We train the conditional prior by minimizing the mean
JS divergence over $\Lambda_{\rm train}$:
\begin{equation}
 \mathcal L_{\rm prior}(\theta)=
 \frac{1}{|\Lambda_{\rm train}|}
 \sum_{\lambda\in\Lambda_{\rm train}}
 \JS
 \bigl(q_\theta(\cdot\mid\lambda),\widehat p_\lambda\bigr).
 \label{eq:prior-loss}
\end{equation}

As illustrated in Figure~\ref{fig:prior-overview},
the shared model can be evaluated at both seen and
unseen parameter values within the training range.
For predictor training at $\lambda^\ast$, we use
$q_\theta(\cdot\mid\lambda^\ast)$ as a fixed statistical
reference.
The prior weights $\theta$ remain fixed, while
optimization updates the predictor weights $\omega$.

\subsection{Graph Quantum Circuit Born Machine}\label{sec:graph}
The Graph QCBM learns conditional distributions over statistical bins
by alternating phase encoding of the system parameter with trainable
Hamiltonian evolution on a fixed graph.
The construction builds on Born-machine probability
representations~\citep{liu2018,benedetti2019}, conditional quantum
circuits, and repeated input encoding~\citep{kiss2022,perez2020}.
It starts from a fixed product reference state estimated from prior
training data and uses one set of trainable weights across parameter
values.

For a \(D\)-dimensional target distribution, 
let the $d$-th dimension contain $M_d$ bins,
giving $M=\prod_{d=1}^{D}M_d$ joint bins.
We construct an undirected graph $G=(V,E)$ in this
discretized space.
Each vertex represents a joint bin
$\mathbf i=(i_1,\ldots,i_D)$, and two vertices
are connected if their bin indices differ by one
along exactly one one grid axis:
\begin{equation}
 V=\prod_{d=1}^{D}\{1,\ldots,M_d\},
 \qquad
 E=\left\{
 \{\mathbf i,\mathbf j\}\subseteq V:
 \sum_{d=1}^{D}|i_d-j_d|=1
 \right\}.
 \label{eq:statistical-graph}
\end{equation}
The graph thus forms a $D$-dimensional
nearest-neighbor grid with open boundaries.

This topology couples adjacent statistical bins,
providing a sparse representation of local
interactions in statistical space.
Repeated graph evolution allows probability
amplitudes to propagate beyond directly connected
bins.

The Graph QCBM associates $M=2^n$ joint bins with
the computational basis states of $n$ qubits.
Writing $i$ for the basis index corresponding
to joint bin $\mathbf i$, the Born rule gives
\begin{equation}
 q_\theta(i\mid\lambda)=
 \left|\langle i\mid\psi_\theta(\lambda)\rangle\right|^2.
 \label{eq:born}
\end{equation}


To construct $\ket{\psi_\theta(\lambda)}$, we initialize
the circuit with a fixed product reference state.
For each dimension $d$, we obtain a reference marginal
$\bar p_d$ by averaging the corresponding empirical
marginals across all training parameter values.
The reference state is defined as
\begin{equation}
 \ket{\psi_{\rm ref}}=\bigotimes_{d=1}^{D}\ket{\psi_d},
 \qquad
 \ket{\psi_d}=\sum_{i_d=1}^{M_d}
 \sqrt{\bar p_d(i_d)}\ket{i_d}.
 \label{eq:reference}
\end{equation}
Its measurement probabilities factorize as
$q_{\rm ref}(i_1,\ldots,i_D)
=\prod_{d=1}^{D}\bar p_d(i_d)$.
This initialization captures the average marginal
statistics and remains fixed across parameter values,
while the subsequent trainable layers learn
dependencies between dimensions and how the joint
distribution changes with the system parameter.

Parameter dependence is introduced through repeated
single-qubit phase rotations~\citep{perez2020,schuld2021}.
For the training interval
$[\lambda_{\min},\lambda_{\max}]$, we define
\begin{align}
 \tau(\lambda)
 &=\frac{2\lambda-\lambda_{\min}-\lambda_{\max}}
 {\lambda_{\max}-\lambda_{\min}},\nonumber\\
 \varphi_{\ell j}(\lambda)
 &=a_{\ell j}+b_{\max}\tanh(v_{\ell j})\,\tau(\lambda),\nonumber\\
 D_\ell(\lambda)
 &=\bigotimes_{j=1}^{n}R_Z\bigl(\varphi_{\ell j}(\lambda)\bigr).
 \label{eq:encoding}
\end{align}
Here $\ell$ and $j$ index the circuit layers and qubits,
respectively, and
$R_Z(\varphi)=\exp(-\mathrm i\varphi Z/2)$.
The trainable parameters $a_{\ell j}$ and $v_{\ell j}$
control the phase offset and its dependence on
$\lambda$. We set $b_{\max}=\pi/2$, so the $\tanh$
parameterization limits the maximum phase change of each
encoding rotation to $\pi$ over the
entire training interval.
This constraint limits rapid phase variation
with the system parameter, particularly when
only a few training conditions are available.

To mix probability amplitudes across statistical bins,
we use continuous time quantum walk evolution on the
graph~\citep{farhi1998,childs2009}.
Each layer has a real symmetric Hamiltonian
defined by the weighted graph adjacency matrix
~\citep[Eq.~(1)]{zimboras2013}:
\begin{equation}
 H_\ell=
 \sum_{\{i,j\}\in E}w_{\ell,ij}
 \bigl(\ket{i}\bra{j}+\ket{j}\bra{i}\bigr),
 \qquad
 U_\ell=\exp(-\mathrm iH_\ell),
 \label{eq:hamiltonian}
\end{equation}
where $w_{\ell,ij}$ is an independently trainable
real coupling for each undirected edge.
The evolution time is absorbed into these couplings.

A phase rotation changes the relative phases of
quantum amplitudes without directly changing their
measurement probabilities.
The subsequent graph evolution mixes these amplitudes,
allowing the encoded parameter to influence the
resulting distribution.
Alternating the two operations through $L$ layers gives
\begin{align}
 U_\theta(\lambda)
 &=U_LD_L(\lambda)\cdots U_1D_1(\lambda),\nonumber\\
 \ket{\psi_\theta(\lambda)}
 &=U_\theta(\lambda)\ket{\psi_{\rm ref}}.
 \label{eq:joint}
\end{align}

\begin{figure}[htbp]
\centering
\vspace{-10mm}
\includegraphics[width=\linewidth]{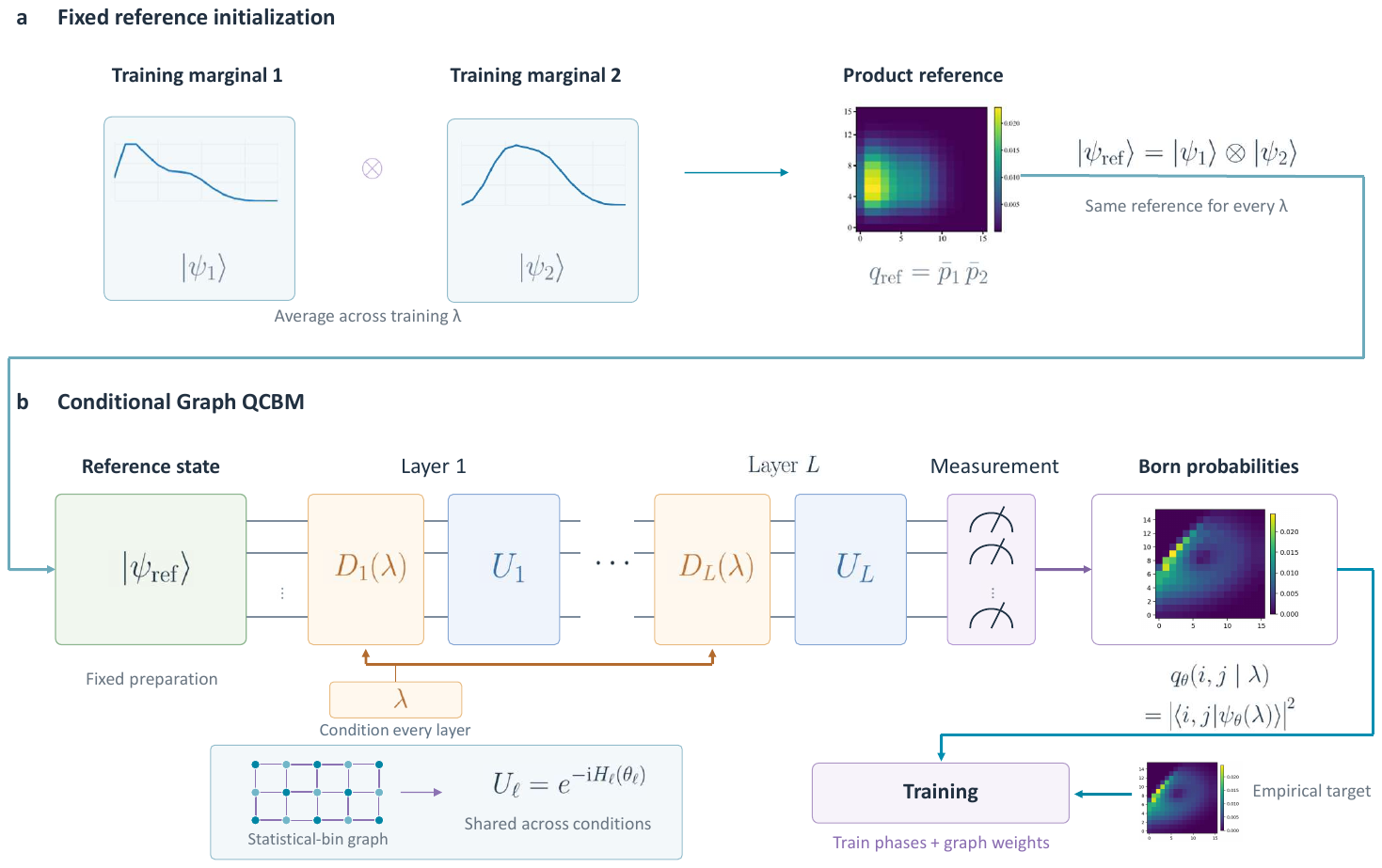}
\caption{Graph QCBM construction.
\textbf{(a)} Fixed product reference state.
\textbf{(b)} Repeated parameter encoding and graph
evolution, followed by measurement to obtain
the conditional joint distribution.}
\label{fig:graph-construction}
\end{figure}
Figure~\ref{fig:graph-construction} illustrates this
construction.
Panel (a) shows the fixed product reference state,
whose initial probabilities are determined by the
reference marginals.
Panel (b) shows how successive phase-encoding
operations $D_\ell(\lambda)$ and graph evolutions
$U_\ell$ transform this state before measurement.
The parameter $\lambda$ enters through the phase
rotations, while the reference state and graph
couplings remain shared across conditions.
Consequently, the trained model can generate
a joint distribution at an unseen parameter value
without requiring empirical statistics from that value.

The trainable weights $\theta$ comprise the graph
couplings $w_{\ell,ij}$ and the phase parameters
$a_{\ell j}$ and $v_{\ell j}$ from all layers.
They are jointly optimized against the empirical
targets at the training parameter values using
Eq.~\eqref{eq:prior-loss}.
For $L$ layers, $|E|$ graph edges, and $n$ qubits,
the total number of trainable weights is
\begin{equation}
 P_{\rm Graph}=L(|E|+2n).
 \label{eq:parameters}
\end{equation}
For our $16\times16$ joint targets, $n=8$,
$|E|=480$, and $L=6$, giving 2,976 trainable
weights, independent of the number of
parameter values queried after training.

\subsection{Photonic Quantum Neural Network}\label{sec:photonic}
The Photonic QNN uses experimentally measured responses
from photonic quantum hardware as nonlinear basis functions~\citep{gan2022}
for conditional distribution learning.
For each optical configuration, we measure the detection
probabilities at different optical phases.
The response associated with detection outcome $k$ is
obtained by normalizing its measured counts:
\begin{equation}
 h_k(\varphi)
 =\frac{N_k(\varphi)}
 {\sum_{k'}N_{k'}(\varphi)},
 \label{eq:photonic-response}
\end{equation}
where $N_k(\varphi)$ denotes the measured counts for
outcome $k$ at phase $\varphi$, and the sum runs over
the retained detection outcomes.

Each response defines a scalar function of the optical
phase. Different optical configurations and detection
outcomes provide a collection of such functions.
The grid bin centers and system parameters are
mapped linearly from their prescribed ranges to
fixed optical phase intervals.
We evaluate the corresponding responses by
interpolating the recorded measurements.
These fixed response functions serve as the basis of
the neural network, while trainable classical coefficients
learn how to combine them.

For the joint statistical targets considered here, let $u_i$ and $v_j$ denote the normalized bin centers
along the first and second axes of the target grid.
We combine the measured optical response functions $h_k(\varphi)$
through trainable classical readouts to obtain
feature vectors
$\boldsymbol\phi_1(u_i),\boldsymbol\phi_2(v_j)
\in\mathbb R^{d_{axis}}$
for the two grid axes, and
$\boldsymbol\chi(\lambda)\in\mathbb R^{d_\lambda}$
for the system parameter.

A trainable classical model combines these
features through individual readouts for each
grid axis and interactions between the axes.
The interaction term is given by
$\boldsymbol\phi_1(u_i)^\top
W_\theta(\boldsymbol\chi(\lambda))
\boldsymbol\phi_2(v_j)$,
where the interaction matrix is
\begin{equation}
 W_\theta(\lambda)
 =
 W_0+\mathcal T_\theta
 \bigl(\boldsymbol\chi(\lambda)\bigr),
\end{equation} 
Here $W_0\in\mathbb R^{d_{\rm axis}\times d_{\rm axis}}$ 
is a trainable matrix, and
$\mathcal T_\theta\in
\mathbb R^{d_\lambda\times d_{\rm axis}\times d_{\rm axis}}$
is a trainable tensor.
Together, they define the
$d_{\rm axis}\times d_{\rm axis}$
interaction matrix $W_\theta(\lambda)$,
which captures parameter-dependent interactions
between the two grid axes.

The interaction term is combined with the
individual readout contributions to obtain
a score $\ell_{ij}(\lambda)$ for each joint bin.
The resulting conditional distribution is
\begin{equation}
 q_\theta(i,j\mid\lambda)
 =\frac{\exp[\ell_{ij}(\lambda)]}
 {\sum_{i',j'}\exp[\ell_{i'j'}(\lambda)]}.
 \label{eq:photonic}
\end{equation}
The parameter features control how the learned
interactions vary with $\lambda$, while the
softmax ensures a normalized joint distribution.

During training, only the classical feature combinations
and interaction parameters are optimized.
The measured photonic responses remain fixed, and the
trainable weights $\theta$ are shared across all
system parameter values.

\subsection{Predictor Regularization}\label{sec:regularization-method}
Regularization incorporates additional constraints into the training objective.
For neural predictors of chaotic dynamics, agreement with short training trajectories alone does not ensure realistic statistics during long autoregressive rollouts~\citep{jiang2023invariant,schiff2024}.
We therefore add a JS penalty that encourages predicted statistics to match the learned prior.
At a system parameter value $\lambda^\ast$, we query the trained
prior to obtain
$q_{\rm prior}=q_\theta(\cdot\mid\lambda^\ast)$.
This distribution serves as a fixed statistical reference during
predictor training.

The classical predictor is trained with a separate objective that combines supervised prediction loss with statistical regularization.
At optimization step $m$, the predictor $G_{\omega_m}$ generates
a finite rollout $\widehat x^{(m)}_{t+1:t+H}$ according to
Eq.~\eqref{eq:rollout}.
From this rollout, we construct $\widehat p_{\omega_m}$,
a finite-sample estimate of the statistics generated by the
current predictor.
We use the same observables, bin grid, and normalization convention
as the prior target in Section~\ref{sec:targets}.

The training objective at step $m$ is
\begin{equation}
 \mathcal L_{\rm predictor}(\omega_m;\lambda^\ast)
 =
 \mathcal L_{\rm data}(\omega_m;\lambda^\ast)
 +\alpha_{\rm reg}
 \JS\bigl(\widehat p_{\omega_m},q_{\rm prior}\bigr),
 \label{eq:predictor}
\end{equation}
where $\mathcal L_{\rm data}$ mean squared
error between predicted and observed next states
and $\alpha_{\rm reg}\geq0$ controls the strength of statistical
regularization. Gradients through the estimated statistics are computed
as described in Appendix~\ref{app:predictor-protocol}.
Optimization updates the predictor weights from $\omega_m$ to
$\omega_{m+1}$, while the prior weights $\theta$ and the reference
distribution $q_{\rm prior}$ remain fixed.

A parameter value excluded from prior training may
still be included in supervised predictor training.
For example, a prior fitted at
$\rho\in\{26,29,32,35,38\}$ can provide
$q_\theta(\cdot\mid28)$ as the statistical reference
for a predictor trained on trajectories at $\rho=28$,
without fitting another prior.

\section{Experiments}\label{sec:experiments}

\subsection{Dynamical systems and prior observables}\label{sec:systems}
We evaluate conditional priors on Lorenz-63, the
Kuramoto--Sivashinsky (KS) equation in one spatial dimension,
and two-dimensional Kolmogorov flow.
These systems provide complementary statistical targets:
attractor occupancy, local derivative statistics, and vorticity
spectra, using the discretization and normalization
defined in Section~\ref{sec:targets}.
The experiments address two questions: how accurately and efficiently the prior represents these statistics, and whether its statistical constraints improve autoregressive prediction.

\paragraph{Lorenz-63.}
Lorenz-63 is a three-mode approximation of the Boussinesq equations
for thermal convection~\citep{lorenz1963}.
The model describes the interaction between convective motion
and temperature variations:
\begin{equation}
 \dot x=\sigma(y-x),\qquad
 \dot y=x(\rho-z)-y,\qquad
 \dot z=xy-\beta z.
 \label{eq:lorenz}
\end{equation}
Here, overdots denote time derivatives. The variable $x$ represents the amplitude of convective motion,
$y$ measures the temperature contrast between ascending and
descending fluid, and $z$ measures the departure of the vertical
temperature profile from linearity~\citep{lorenz1963}.
The parameters $\sigma$, $\rho$, and $\beta$ specify the Prandtl number, the Rayleigh number relative to its critical value, and a geometric coefficient, respectively.
We fix $\sigma=10$ and $\beta=8/3$ and vary the thermal driving through $\lambda=\rho\in[26,38]$ in both the Graph QCBM and Photonic QNN.

The prior represents the joint occupancy distribution of
$r=\sqrt{x^2+y^2}$ and $z$ on a $16\times16$ bin grid.
Empirical targets are estimated from trajectory samples.
Both $r$ and $z$ are invariant under the Lorenz symmetry
$(x,y,z)\mapsto(-x,-y,z)$.
The joint distribution captures the dependence between the radial amplitude and deviations of the vertical temperature profile from linearity.

\paragraph{Kuramoto--Sivashinsky.}
The KS equation describes weakly nonlinear instabilities in extended systems, including phase dynamics in reaction-diffusion media~\citep{kuramoto1976}.
It also arises as an approximation to Navier--Stokes flow at long wavelengths along the free surface of a thin liquid film~\citep{ueno2015}.
This connection makes KS a reduced model of spatial instability. The scalar field represents a reduced amplitude whose physical meaning depends on the derivation.
We use the periodic equation
\begin{equation}
 \partial_tu+u\partial_xu+\partial_{xx}u+\nu\partial_{xxxx}u=0,
 \qquad \nu>0.
 \label{eq:ks}
\end{equation}
The second derivative amplifies perturbations at long wavelengths, while the fourth derivative damps short wavelengths with strength $\nu$.
Nonlinear advection transfers energy between spatial scales.

For a periodic domain of length $L_x$, multiplying Eq.~\eqref{eq:ks} by $u$ and integrating by parts gives
\begin{equation}
 \frac{\mathrm d}{\mathrm dt}\frac12\int_0^{L_x}u^2\,\mathrm dx
 =\int_0^{L_x}u_x^2\,\mathrm dx
 -\nu\int_0^{L_x}u_{xx}^2\,\mathrm dx.
 \label{eq:ks-energy}
\end{equation}
The squared derivatives $u_x^2$ and $u_{xx}^2$ therefore determine
the production and dissipation of this quadratic energy.

Motivated by this balance, we use the joint distribution of
$u_x=\partial_xu$ and $u_{xx}=\partial_{xx}u$ as the prior target.
Its second moments characterize the mean production and
dissipation terms, while its joint structure captures dependence
between local gradient and curvature.
The finite bin representation provides a discretized approximation
to these statistics.
Pooling over spatial locations makes the underlying statistic
invariant to translations of the periodic field, while discarding
the spatial arrangement of the derivative values.
The KS experiments on conditional priors and regularization use $L_x=64\pi$ and $\lambda=\nu\in[0.6,1.2]$.

\paragraph{Kolmogorov flow.}
We consider two-dimensional Kolmogorov flow, an
incompressible Navier--Stokes system driven by
a sinusoidal body force on a periodic
domain~\citep{lucas2015}.
The governing equations, including linear drag, are:
\begin{equation}
 \begin{aligned}
 \partial_t\mathbf u+(\mathbf u\cdot\nabla)\mathbf u
 &=
 -\nabla p+Re^{-1}\nabla^2\mathbf u
 +A\sin(k_fy)\mathbf e_x-0.1\,\mathbf u,
 \\
 \nabla\cdot\mathbf u&=0.
 \end{aligned}
 \label{eq:kf}
\end{equation}
Here $Re$ is the Reynolds number, $A$ is the forcing amplitude,
$k_f$ is the forcing wavenumber, and $\mathbf e_x$ is the
streamwise unit vector. The term $-0.1\,\mathbf u$ represents linear drag
in addition to viscous diffusion.

The classical predictor forecasts the scalar vorticity field
$\zeta=\partial_xu_y-\partial_yu_x$. Taking the curl of Eq.~\eqref{eq:kf} gives
\begin{equation}
 \partial_t\zeta+\mathbf u\cdot\nabla\zeta
 =
 Re^{-1}\nabla^2\zeta
 -Ak_f\cos(k_fy)-0.1\,\zeta.
 \label{eq:kf-vorticity}
\end{equation}
The forcing injects energy at a prescribed spatial scale,
while nonlinear advection redistributes energy among
Fourier modes and viscosity dissipates it.
Varying $Re$ changes the balance between these processes
and the resulting long-time flow statistics~\citep{lucas2015}.

For the conditional Graph QCBM experiments, we vary
$Re\in[400,1000]$ at fixed $A=1$ and $k_f=8$.
The statistical target is the normalized, time-averaged
vorticity power spectrum, represented on a
$16\times16$ grid of wavenumbers.
For each trajectory, we average the squared Fourier
amplitudes of vorticity over time, combine contributions
from wavenumbers with the same absolute components,
and aggregate the resulting power into the statistical
grid. The outermost bins also collect contributions
from higher wavenumbers.


\subsection{Conditional prior learning}\label{sec:prior-results}

We evaluate two implementations of conditional prior learning: the Graph QCBM and the Photonic QNN. For the Graph QCBM, we examine training coverage, generalization accuracy, parameter efficiency, and estimated implementation cost on gate-based quantum hardware. The Photonic QNN is then evaluated using experimentally measured optical responses.

\subsubsection{Training Coverage and Generalization}\label{sec:coverage}
We vary the number $N_{\rm train}$  of training conditions to determine how much parameter coverage the prior needs to generalize.
Each study fixes the architecture, statistical grid, and unseen
evaluation values, with a prescribed accuracy threshold of
$\mathrm{JS}\leq5\times10^{-3}$.
Lorenz-63 and KS use
$N_{\rm train}\in\{2,3,5,9,17\}$, while Kolmogorov flow uses
$N_{\rm train}\in\{2,3,5\}$.
Table~\ref{tab:coverage} reports the maximum JS across all unseen
evaluation values and three training runs for each training set size.
Appendix~\ref{app:coverage-grids} specifies the exact parameter grids
and training protocols.

\begin{table}[h!]
\centering\small
\begin{tabular}{lrrrr}
\toprule
System & Unseen conditions & JS ($N_{\rm train}=2$) $\downarrow$ & JS ($N_{\rm train}=3$) $\downarrow$ & JS ($N_{\rm train}=5$) $\downarrow$ \\
 & & $(\times10^{-3})$ & $(\times10^{-3})$ & $(\times10^{-3})$ \\
\midrule
Lorenz-63 & 96 & 121.481 & 23.426 & 2.804 \\
KS & 16 & 102.184 & 22.316 & 1.283 \\
Kolmogorov flow & 16 & 5.222 & 0.316 & 0.117 \\
\bottomrule
\end{tabular}
\caption{Maximum unseen JS in simulation over three seeds and all evaluation conditions for $m$ training conditions.}
\label{tab:coverage}
\end{table}

Table~\ref{tab:coverage} shows that, with five training conditions, the maximum unseen JS divergence is below the prescribed threshold of $5\times10^{-3}$for all three systems. Among the tested training set sizes, three conditions are sufficient for Kolmogorov flow, whereas Lorenz-63 and KS require five. We therefore use five training conditions in the subsequent Graph QCBM experiments.
Figure~\ref{fig:condition-responses} in Appendix~\ref{app:coverage-results} shows that denser Lorenz training coverage further improves agreement with unseen empirical targets.

\begin{figure}[h!]
\centering
\includegraphics[width=0.94\linewidth]{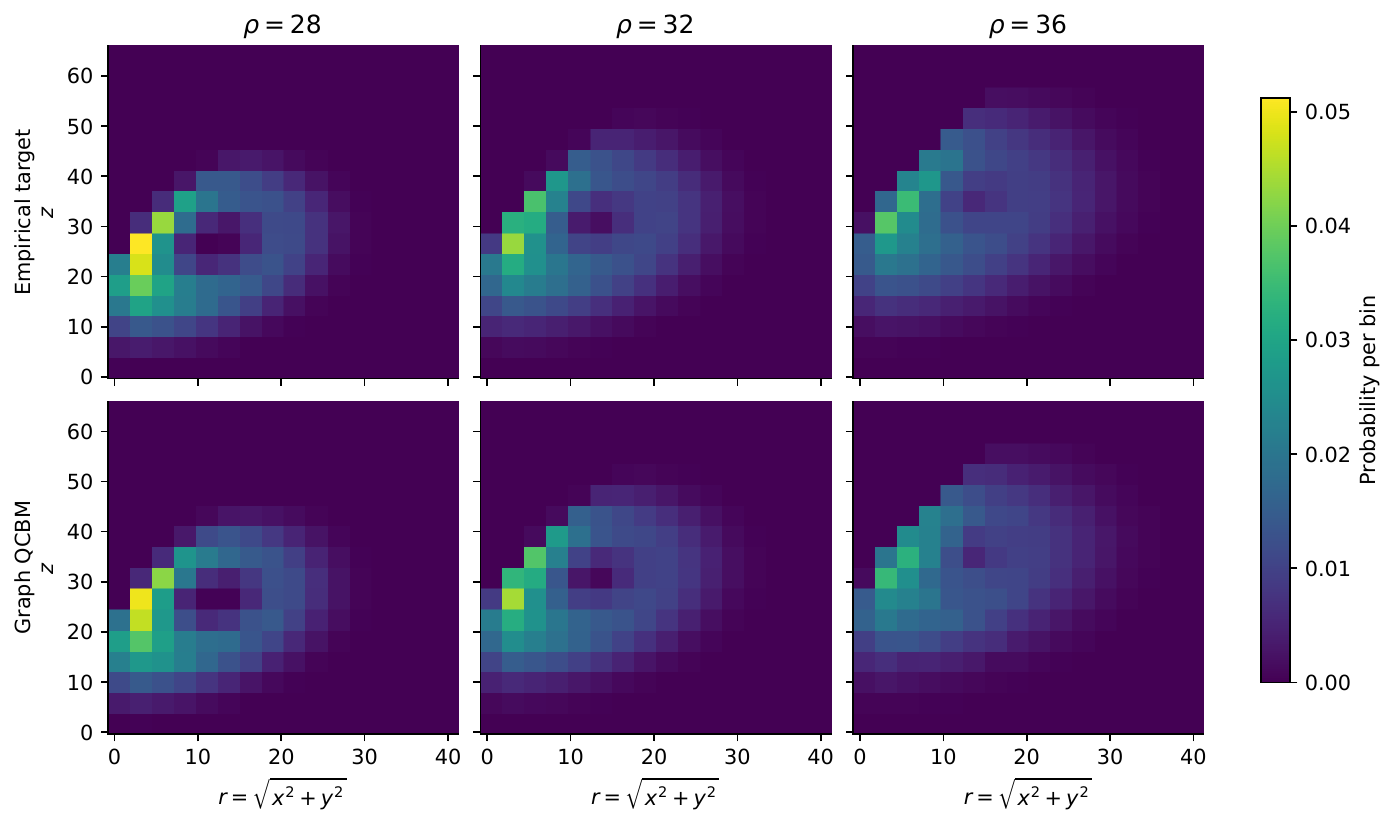}
\caption{Lorenz-63 joint distributions at
$\rho=28,32,36$.
Empirical targets (top) and Graph QCBM predictions
(bottom).}
\label{fig:interpolation}
\end{figure}

Figure~\ref{fig:interpolation} compares the empirical Lorenz joint distributions with the Graph QCBM predictions at three parameter values. Among the three values shown, \(\rho=32\) was included in prior training, whereas \(\rho=28\) and \(\rho=36\) were unseen.
As thermal driving increases, the high-probability
region shifts toward larger radial and vertical
amplitudes.
The learned distributions reproduce this change
at both seen and unseen conditions, showing how
the shared prior captures the dependence of the
joint statistics on the system parameter.

\subsubsection{Parameter Efficiency and Accuracy}\label{sec:interpolation}\label{sec:architecture-results}\label{sec:resources}
The Lorenz evaluation grid contains five training values and
96 unseen values, giving 101 system parameter values in total.
The QNN used in QIML has 192 trainable weights and does not take
the system parameter as input~\citep{wang2026}.
A separate QNN at each of the 101 values would therefore
require 19,392 trainable weights.
One conditional Graph QCBM uses 2,976 shared weights,
reducing this projected total by 84.7\%.

\begin{table}[h!]
\centering\small
\begin{tabular}{lcc}
\toprule
Comparison & QIML QNN~\citep{wang2026} & Graph QCBM (ours) \\
\midrule
\shortstack[l]{Trainable weights and storage\\(101 conditions)} &
\shortstack{19,392 weights\\151.50 KiB (float64)} &
\shortstack{2,976 weights\\23.25 KiB (float64)} \\[4pt]
Test JS at $\rho=28$ $(\times10^{-3})$ $\downarrow$
& $71.0473$ & $0.59282$ \\
\bottomrule
\end{tabular}
\caption{Parameter efficiency and simulated prior accuracy
for Lorenz-63.
The projected QIML QNN cost assumes a separate model
at each parameter value.}
\label{tab:qnn-sharing-cost}
\end{table}

For the accuracy comparison, we train the QIML reference
QNN with eight qubits and eight layers, retaining its
original circuit and optimization settings~\citep{wang2026}.
Both models are evaluated against the same Lorenz
$p(r,z)$ target on a $16\times16$ bin grid.
As shown in Table~\ref{tab:qnn-sharing-cost},
the Graph QCBM achieves a test JS of
$0.59282\times10^{-3}$ at $\rho=28$,
compared with $71.0473\times10^{-3}$ for the QIML QNN,
corresponding to a 99.2\% reduction.

We next compare four conditional architectures with similar
numbers of trainable weights.
Each model takes the system parameter as input.
The conditional QNN retains the QIML rotations and
controlled-NOT (CNOT) gates~\citep{wang2026}, incorporates the
Graph QCBM phase encoding, and uses additional circuit layers
to reach 2,976 trainable weights.
A conditional multilayer perceptron (MLP) and a quantum walk
with a trainable coin operator~\citep{chang2026quantumwalks} provide classical and quantum comparisons.
In the coin model, an auxiliary quantum state controls shifts
between statistical bins, and trainable rotations of that
state determine the resulting distribution.
The three quantum models use the same fixed product reference
state over the statistical bins.

All four models minimize mean JS on the same five training
targets.
They use the same optimization schedule: 1,000 Adam updates
with a learning rate of $0.002$, repeated over five seeds.
We select checkpoints using the same nine validation values
and evaluate test targets estimated from independent
trajectories at five seen and 96 unseen parameter values.

\begin{table}[h!]
\centering\small
\begin{tabular}{lrrr}
\toprule
Model & Trainable weights & Seen JS $\downarrow$ & Unseen JS $\downarrow$ \\
 & & $(\times10^{-3})$ & $(\times10^{-3})$ \\
\midrule
Graph QCBM & 2,976 & $\mathbf{0.3038}\pm0.0186$ & $\mathbf{0.5336}\pm0.0181$ \\
Conditional QNN (QIML circuit) & 2,976 & $0.3121\pm0.0065$ & $0.5624\pm0.0075$ \\
Conditional MLP & 2,963 & $1.3734\pm0.5603$ & $1.2385\pm0.4363$ \\
Conditional coin quantum walk & 2,976 & $1.2378\pm0.1069$ & $1.3202\pm0.1097$ \\
\bottomrule
\end{tabular}
\caption{Conditional prior accuracy for Lorenz-63 at comparable numbers of trainable weights.
Values give mean JS $\pm$ sample standard deviation across five seeds, after averaging over five seen or 96 unseen test conditions.}
\label{tab:interpolation}
\end{table}

Table~\ref{tab:interpolation} shows that the Graph QCBM
achieves the lowest mean JS at both seen and unseen
parameter values among the four tested models.
Its mean unseen JS is 56.9\% lower than that of the conditional MLP. Relative to the conditional QNN, mean JS decreases by 2.7\% at seen conditions and 5.1\% at unseen conditions.
These results support the Graph QCBM's accuracy at both seen and unseen conditions under a common training budget.
Appendix~\ref{app:architecture-results} further investigates the tradeoff between graph size and distribution accuracy.

\subsubsection{Hardware Feasibility}\label{sec:photonic-results}
Executing the Graph QCBM on a quantum processor requires
decomposing its graph evolution into native gates.
Using the IBM Torino calibration dated 26 February 2025,
we compile an approximate superconducting quantum circuit at $\rho=30.5$
for eight physical qubits, including reference state preparation.
The compiled circuit contains 28,462 controlled-$Z$ (CZ)
gates and has depth 58,871.
We summarize the calibrated gate errors using the heuristic score
\begin{equation}
 F_{\rm gates}
 =
 \prod_{g\in\mathcal C}(1-\epsilon_g),
 \label{eq:gate-product}
\end{equation}
where $\mathcal C$ is the compiled gate sequence and
$\epsilon_g$ is the calibrated error reported for gate $g$.
If the gate errors were independent failure
probabilities, $F_{\rm gates}$ would represent
the probability of completing the circuit
without a gate error. However, it does not capture coherent or
correlated errors, idle decoherence, or readout
noise.
Therefore, this score provides a rough measure
of cumulative noise exposure,
The value $F_{\rm gates}=4.66\times10^{-46}$ highlights the challenges of implementing
the Graph QCBM for this compiled realization, 

The large gate count and circuit depth motivate exploring an alternative approach that does not require implementing the graph evolution as a deep gate-based quantum circuit. The Photonic QNN instead uses experimentally measured optical responses as fixed features, combined with trainable classical components to learn conditional prior distributions, as described in Section~\ref{sec:photonic}.
For Lorenz-63, the Photonic QNN uses 17 training values
in $\rho\in[26,38]$, since training at five values
does not meet the prescribed JS criterion.
For Kolmogorov flow, five training values in
$Re\in[400,1000]$ are sufficient.
The Graph QCBM uses five training values for both systems.

\begin{figure}[h!]
\centering
\begin{subfigure}{\linewidth}
\centering
\includegraphics[width=\linewidth]{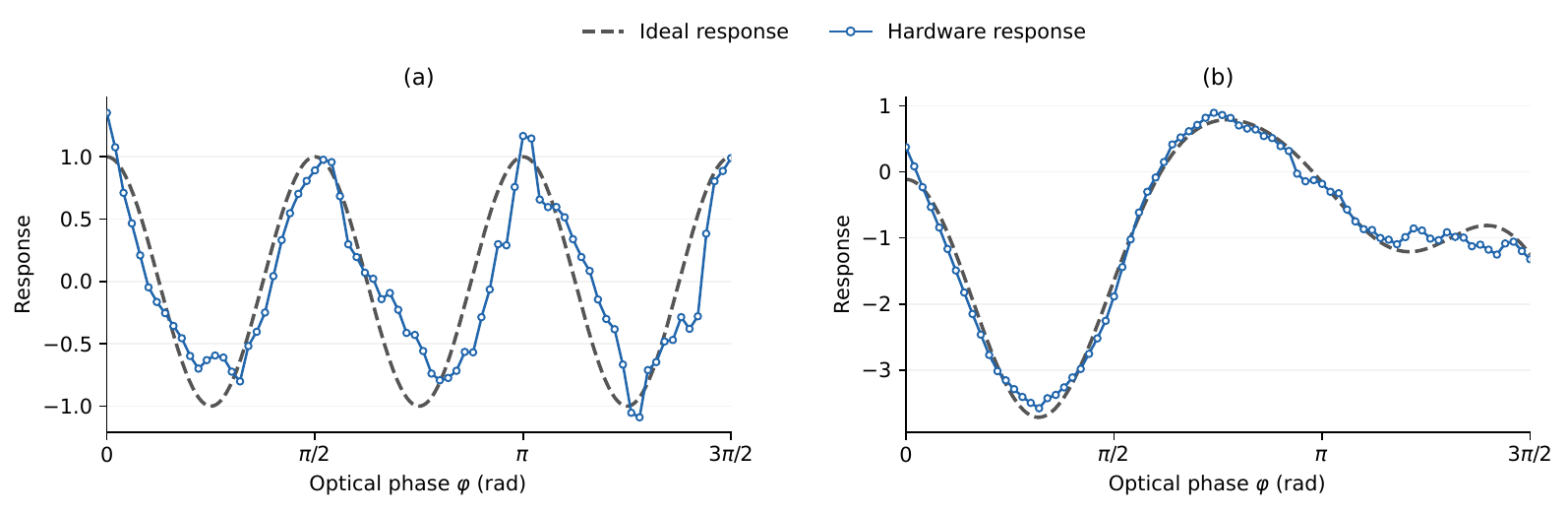}
\phantomcaption
\label{fig:photonic-readout}
\end{subfigure}
\par\medskip
\begin{subfigure}{\linewidth}
\centering
\includegraphics[width=\linewidth]{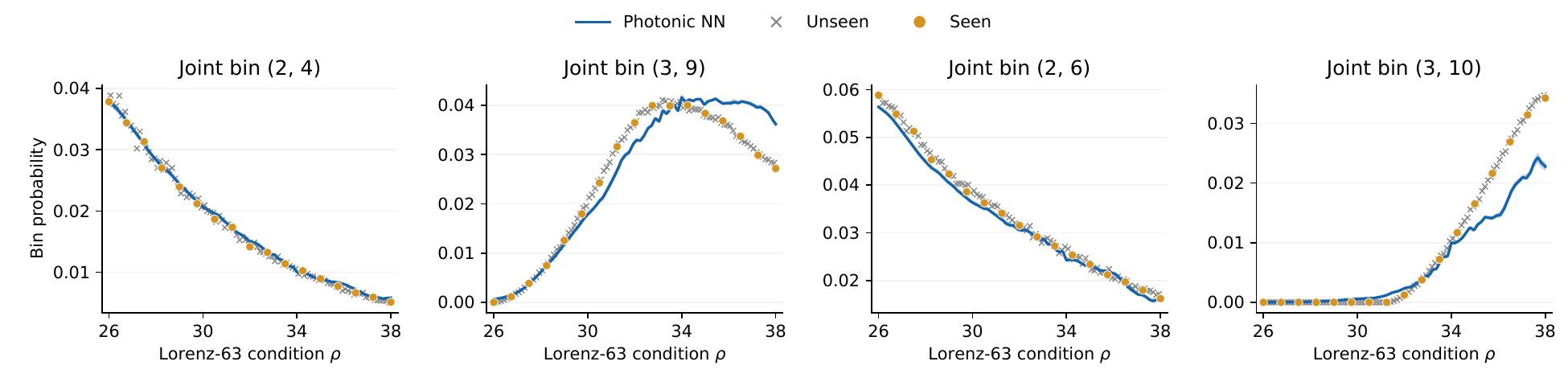}
\phantomcaption
\label{fig:photonic-bin-responses}
\end{subfigure}

\caption{Photonic function reconstruction and conditional
prior predictions.
\textbf{Top.} Ideal functions (dashed) contructed by ideal responses and readouts
from measured optical responses (blue):
(a) $\cos(4\varphi)$ and
(b) a feature function from the prior fitted at $\rho=28$.
Circles mark the measured phase values.
\textbf{Bottom.} Lorenz bin probabilities predicted by
the Photonic QNN trained at 17 conditions.
Orange circles and gray crosses
denote training and unseen empirical targets, respectively.}
\label{fig:photonic-lorenz-responses}
\end{figure}

Table~\ref{tab:photonic-conditional} compares the
conditional prior accuracy of the two implementations.
The Photonic QNN achieves similar errors at seen and
unseen parameter values in both systems.
Its mean unseen JS is $4.5761\times10^{-3}$
for Lorenz-63 and $2.585\times10^{-3}$
for Kolmogorov flow, with a standard deviation of
$0.238\times10^{-3}$ for the latter.
Although these errors are higher than those of
the Graph QCBM in ideal simulation, the results
demonstrate that measured photonic responses can
support conditional prior learning across
different parameter values.
\begin{table}[h!]
\centering
\small
\setlength{\tabcolsep}{5pt}
\renewcommand{\arraystretch}{1.15}
\begin{tabular}{@{}l@{\hspace{2em}}crrrr@{}}
\toprule
Model & Implementation & Weights & Training conditions
& Seen JS $\downarrow$ & Unseen JS $\downarrow$ \\
 & & & & $(\times10^{-3})$ & $(\times10^{-3})$ \\
\midrule
\multicolumn{6}{@{}l}{\textbf{Lorenz-63}} \\
Graph QCBM & Simulation & 2,976 & 5 & 0.3038 & 0.5336 \\
Photonic QNN & Hardware & 1,862 & 17 & 4.569 & 4.5761 \\
\midrule
\multicolumn{6}{@{}l}{\textbf{Kolmogorov flow}} \\
Graph QCBM & Simulation & 2,976 & 5 & 0.0902 & 0.08954 \\
Photonic QNN & Hardware & 1,234 & 5 & 2.585 & 2.585 \\
\bottomrule
\end{tabular}
\caption{Conditional prior accuracy in ideal simulation
and using photonic hardware measurements.
JS values are averaged over evaluation conditions
and three training runs.}
\label{tab:photonic-conditional}
\end{table}
Figure~\ref{fig:photonic-lorenz-responses} further
illustrates the effectiveness of this construction.
The upper panels show that classical readouts
of measured optical responses reproduce the
overall behavior of the ideal scalar functions.
Some differences in amplitude and shape remain,
potentially arising from photonic hardware noise and
imperfections in the measured responses.
The lower panels show that the learned prior
captures the changes in Lorenz bin probabilities
across both training and unseen parameter values,
although deviations remain in some bins toward
the upper end of the parameter range.
Together, these results show that the Photonic QNN
can learn parameter-dependent statistical distributions
from measured photonic responses on quantum hardware.
Further details of the optical response processing
and model dimensions are provided in
Appendix~\ref{app:photonic-protocol}.

Having examined the accuracy and hardware feasibility
of conditional prior learning, we next evaluate
how the trained priors affect the training of classical predictor.

\subsection{Effect of Prior Regularization}\label{sec:reuse-results}
Regularization with the trained Graph QCBM priors reduces
test trajectory mean squared error (MSE) in all three
evaluated systems.
For Kolmogorov flow and KS, we use the classical predictor architectures and data objectives from QIML~\citep{wang2026}.
The \emph{Without prior} predictor uses the data objective
alone, with $\alpha_{\rm reg}=0$.
The \emph{Quantum prior} predictor adds the JS penalty in
Eq.~\eqref{eq:predictor}, with $\alpha_{\rm reg}=0.01$.
The prior weights remain fixed throughout predictor training.
Each predictor is trained on trajectories at the parameter
value being evaluated.
Lorenz $\rho=28$ and KS $\nu=1$ are excluded from prior
fitting, whereas Kolmogorov flow $Re=1000$ is included.
Thus, seen and unseen refer to prior training.
Table~\ref{tab:reuse} reports the resulting test trajectory MSE. The largest reduction occurs for Kolmogorov flow
(66.0\%), followed by Lorenz-63 (32.9\%),
while the improvement for KS is more modest (1.42\%).
The following comparisons examine these differences
in trajectory and statistical behavior.

\begin{table}[h!]
\centering
\small
\setlength{\tabcolsep}{5pt}
\renewcommand{\arraystretch}{1.18}

\begin{tabular}{@{}lrrr@{}}
\toprule
System and condition
& Without prior MSE $\downarrow$
& Quantum prior MSE $\downarrow$
& Reduction \\
\midrule
Lorenz ($\rho=28$, unseen) & 74.496 & 50.016 & 32.9\% \\
Kolmogorov flow ($Re=1000$, seen) & 196.34 & 66.74 & 66.0\% \\
KS ($\nu=1$, unseen) & 3.41291 & 3.36437 & 1.42\% \\
\bottomrule
\end{tabular}

\caption{Test trajectory MSE averaged over 500 steps for Lorenz and KS
and 318 steps for Kolmogorov flow.
Quantum prior uses $\alpha_{\rm reg}=0.01$;
without prior uses $\alpha_{\rm reg}=0$.}
\label{tab:reuse}
\end{table}

For Lorenz-63, the conditional prior improves predictions at
$\rho=28$, which was excluded from prior training and lies
between the training values $\rho=26$ and $\rho=29$.
Both predictors share initialization, training data, and batch order.
Prior regularization reduces MSE over 500 prediction steps by
32.9\% (Table~\ref{tab:reuse}).

\begin{figure}[h!]
\centering
\begin{subfigure}{\linewidth}
\centering
\includegraphics[width=\linewidth]{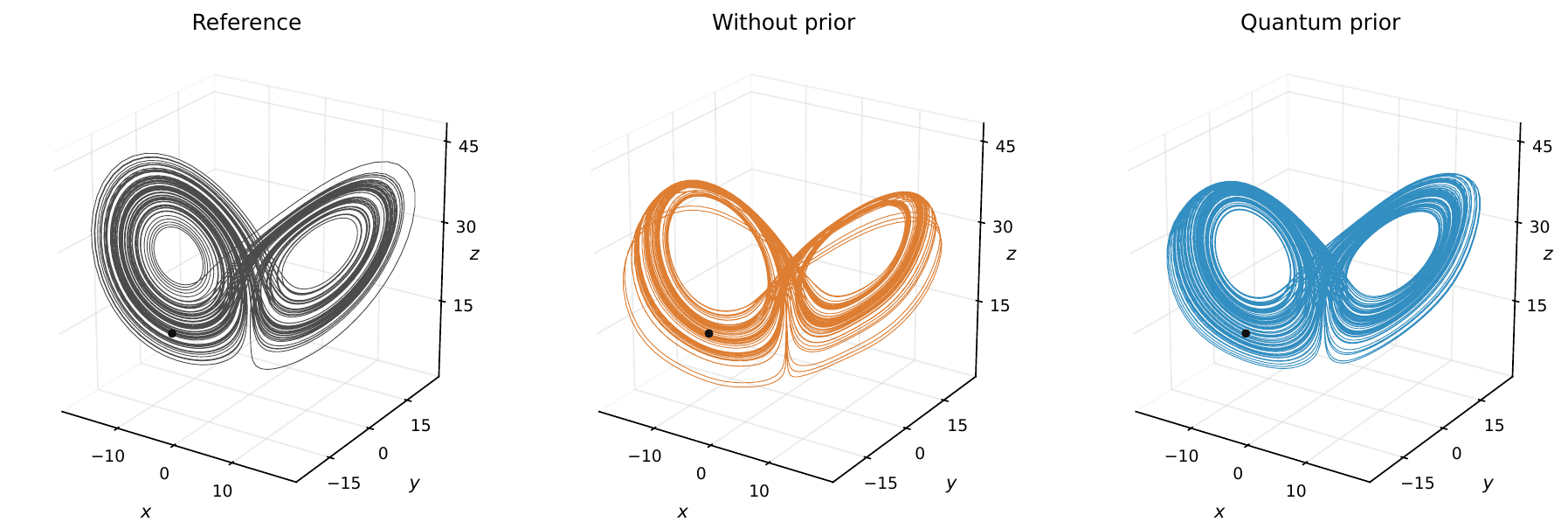}
\phantomcaption
\label{fig:lorenz-full-trajectory}
\end{subfigure}
\par\medskip
\begin{subfigure}{\linewidth}
\centering
\includegraphics[width=\linewidth]{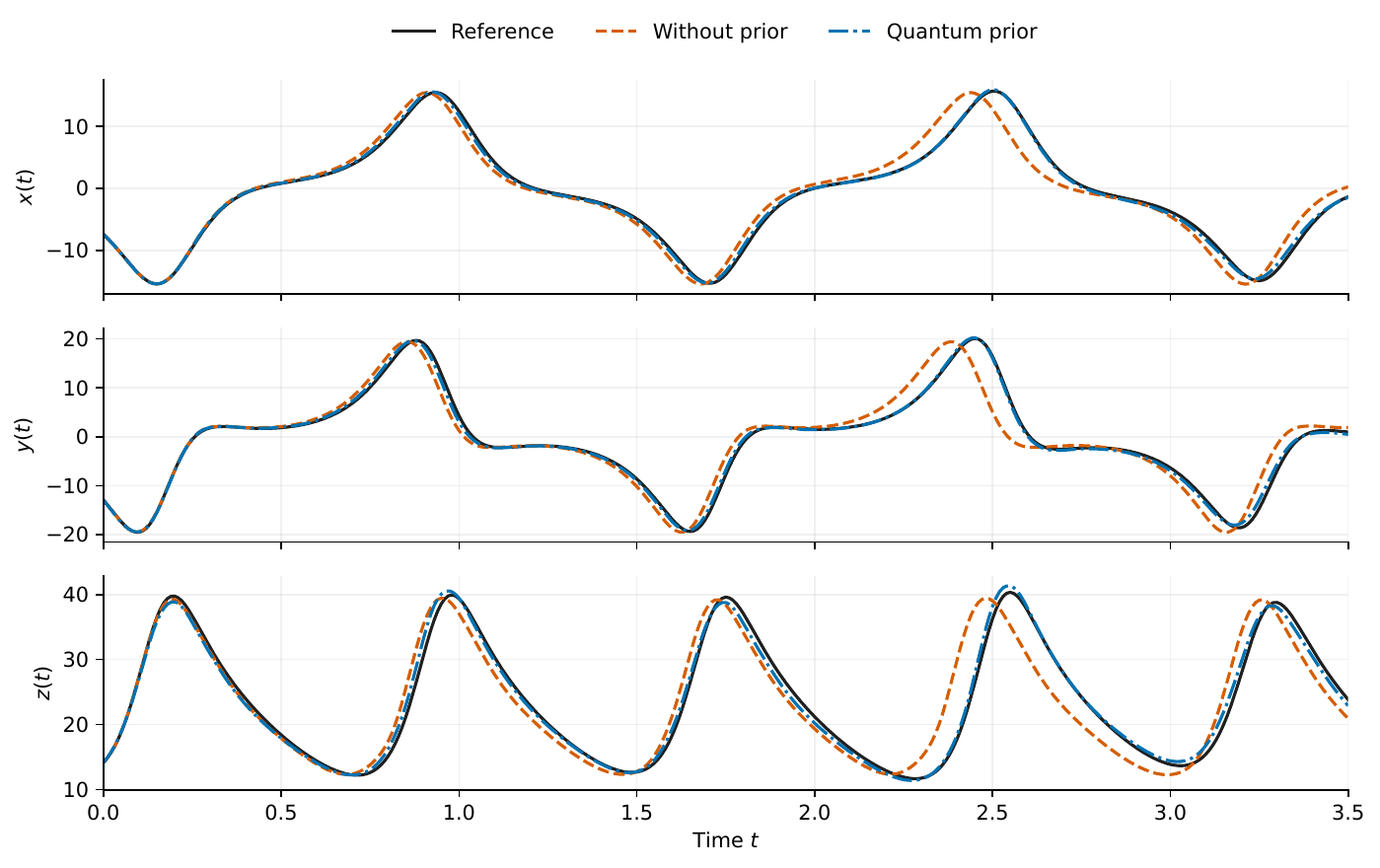}
\phantomcaption
\label{fig:lorenz-state-evolution}
\end{subfigure}
\caption{Lorenz-63 forecasts at $\rho=28$,
unseen during prior training.
\textbf{A (top).} Three-dimensional trajectories
over $0\leq t\leq100$.
\textbf{B (bottom).} Coordinate evolution over
$0\leq t\leq3.5$.
All predictions use the same initial state
with $\Delta t=0.01$.}
\label{fig:lorenz-regularization}
\end{figure}

To assess statistics over longer horizons, we separately evaluate
a rollout of 10,000 prediction steps.
Over the final 5,000 steps, the JS divergence between
the predicted and reference $(r,z)$ distributions decreases
by 49.5\%, from $48.46\times10^{-3}$ to
$24.49\times10^{-3}$.
Figure~\ref{fig:lorenz-regularization} provides two
views of the predictions from a common initial state.
The upper panel compares the complete trajectories
in the three-dimensional Lorenz state space,
while the lower panel shows the early evolution
of the individual state variables.
The full trajectories illustrate the regions of
the attractor visited by the predictors, whereas
the coordinate trajectories allow a closer comparison
of their individual evolution.
Although long-term pointwise agreement is difficult
to maintain in chaotic dynamics, the regularized
predictor achieves lower forecast MSE.
Together, these results show how the learned prior
can improve prediction at a parameter value
excluded from prior training.

At $Re=1000$, the Kolmogorov flow predictors share initialization, training data, batch size, and update count.
The prior represents normalized vorticity power across spatial scales and directions.
Regularization with this prior reduces test vorticity MSE by 66.0\% over 318 forecast steps.

\begin{figure}[h!]
\centering
\begin{subfigure}{\linewidth}
\centering
\includegraphics[width=0.82\linewidth,height=0.40\textheight,keepaspectratio]{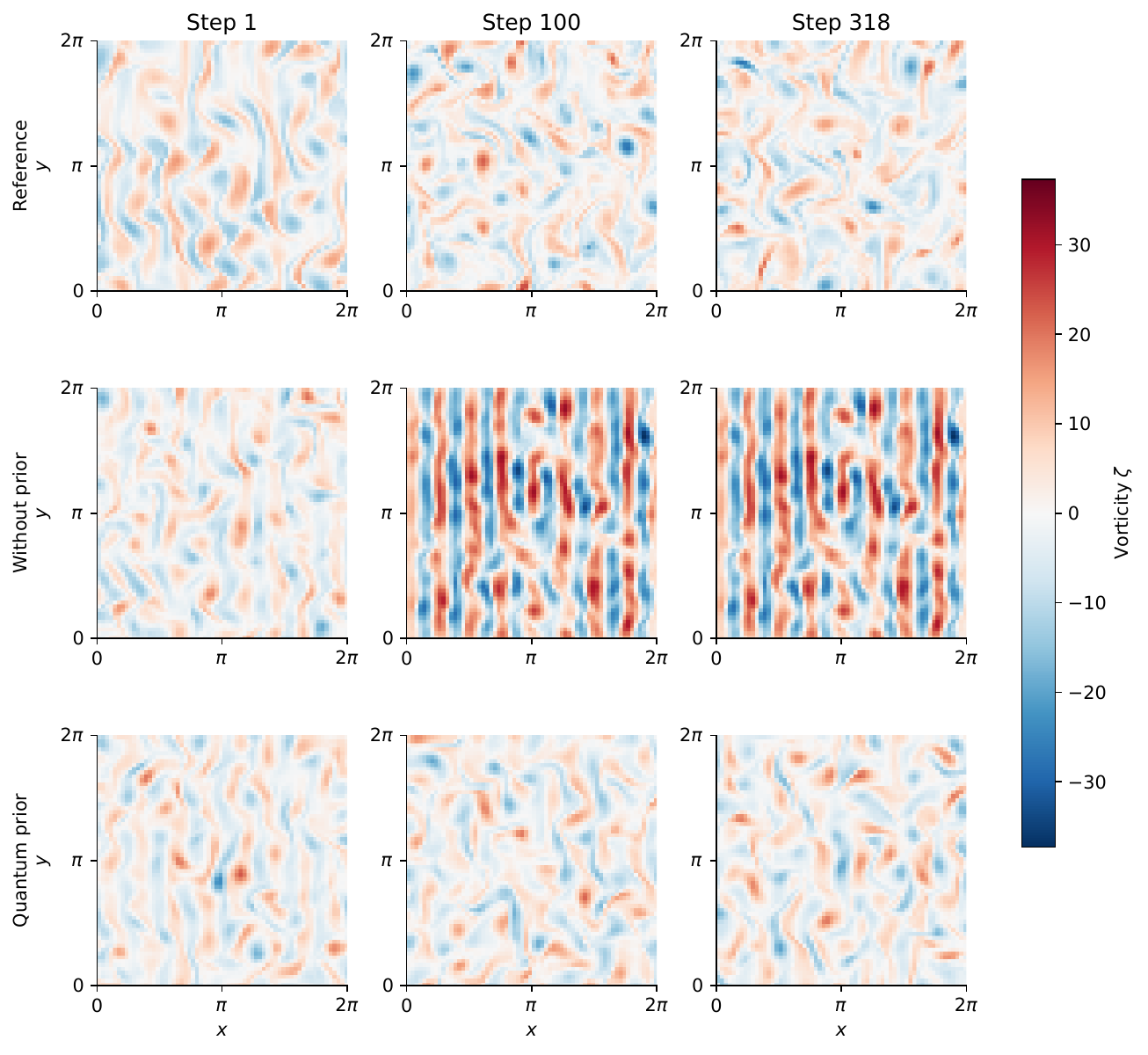}
\phantomcaption
\label{fig:kf-snapshots}
\end{subfigure}
\par\medskip
\begin{subfigure}{\linewidth}
\centering
\includegraphics[width=\linewidth,height=0.33\textheight,keepaspectratio]{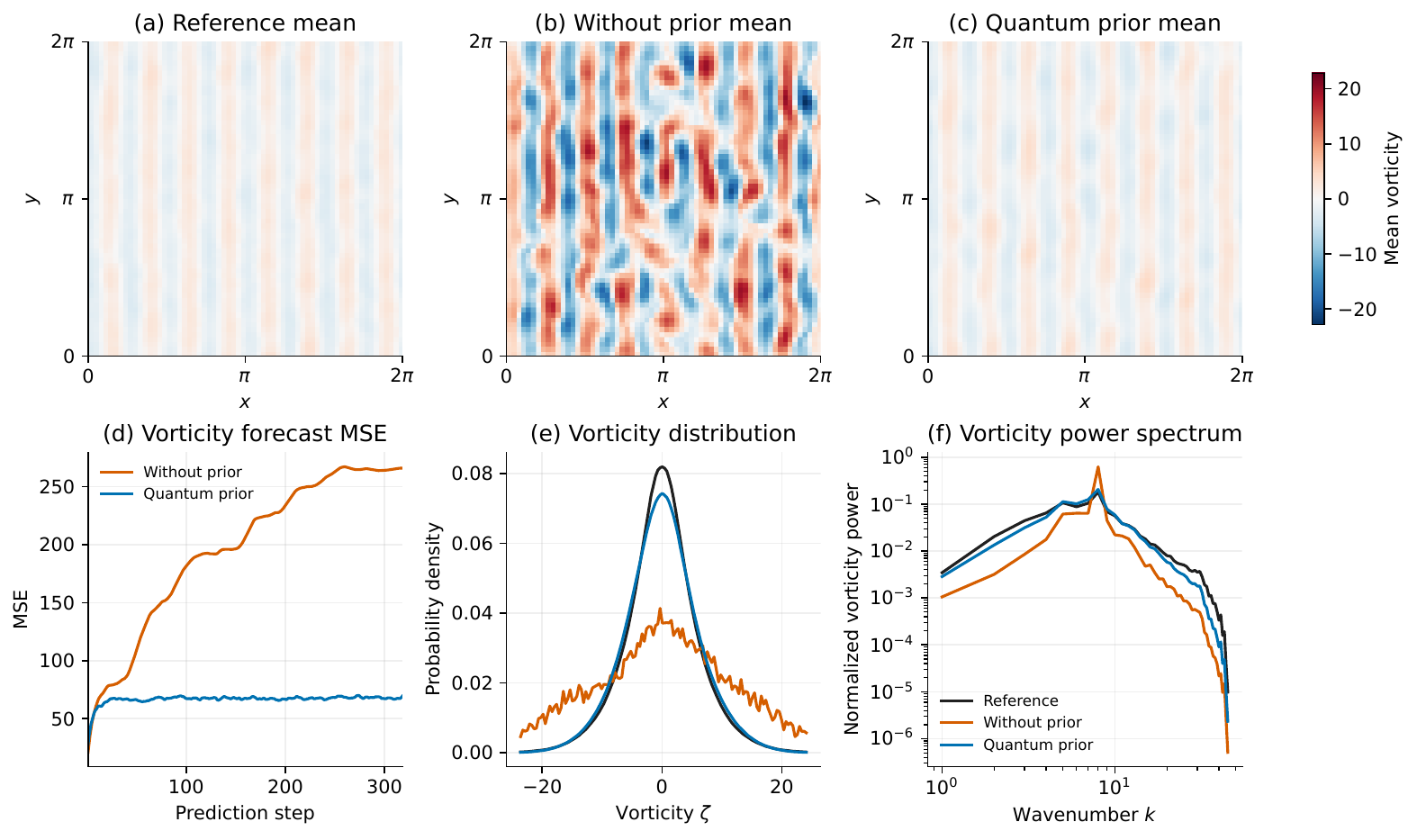}
\phantomcaption
\label{fig:kf-statistics}
\end{subfigure}
\caption{Kolmogorov flow forecasts at $Re=1000$.
\textbf{A (top).} Reference, unregularized, and
regularized vorticity fields at prediction steps
1, 100, and 318, shown on a common color scale.
\textbf{B (bottom).} Time-averaged vorticity fields
(a--c), forecast MSE (d), vorticity distributions
(e), and normalized radial power spectra (f),
evaluated over 20 test trajectories.}
\label{fig:kf-regularization}
\end{figure}

Figure~\ref{fig:kf-regularization} compares the
predicted Kolmogorov flow fields and their statistical properties.
Panel A shows instantaneous vorticity fields at
three forecast steps, while panel B compares the
time-averaged fields, forecast errors, vorticity
distributions, and power spectra.
At the first forecast step, both predictors retain spatial
structure similar to the reference.
At forecast steps 100 and 318, the predictor without the
prior produces a band structure that differs markedly from
the reference vorticity patterns.
These bands remain nearly unchanged between the two
snapshots, suggesting a loss of temporal variability.
The predictor regularized with the quantum prior better
preserves the characteristic spatial structure and amplitude
range of the reference fields, while producing distinct
vorticity patterns at the two prediction steps.

Averaging the vorticity fields over time and the 20 test
trajectories reveals alternating bands, consistent with
the mean shear sustained by the sinusoidal
forcing~\citep{lucas2015}.
The regularized predictor matches the reference
band structure and amplitude more closely.
The predictor without the prior produces stronger,
spatially distorted bands.
For these averaged fields, the spatial MSE relative
to the reference mean field decreases from $96.684$
to $0.615$, a 99.4\% reduction.

The probability density function (PDF) and spectrum show corresponding improvements in vorticity amplitudes and spatial scales.
Regularization brings the broader vorticity distribution of the baseline closer to the reference, reducing JS divergence from $90.17\times10^{-3}$ to $1.007\times10^{-3}$.
It also reduces the excessive concentration of power near the forcing wavenumber and better reproduces the distribution across scales.
The improvements in the mean vorticity field and the vorticity PDF extend beyond the spectral statistic used for regularization, showing better preservation of the flow's persistent statistical structure.

For KS at $\nu=1$, regularization improves the joint
derivative statistics over prediction steps 251--500.
The JS divergence between the predicted and reference
distributions of $(u_x,u_{xx})$ decreases by 31.2\%,
from $6.467\times10^{-3}$ to $4.449\times10^{-3}$.
The corresponding reduction in trajectory MSE is more
modest, at 1.42\%.
This smaller gain compared with Lorenz-63 and Kolmogorov
flow likely reflects, in part, the baseline's already good
agreement with the KS derivative statistics, leaving less
room for improvement through statistical regularization.

Across the three systems, conditional priors improve autoregressive prediction. 
For Lorenz-63 and KS, these improvements are achieved at
parameter values not used to train the prior.
The Kolmogorov flow results additionally show closer
agreement with reference mean fields, vorticity distributions,
and power spectra over the evaluated forecast window.

\section{Discussion}\label{sec:discussion}
The results show that a single conditional quantum
prior can capture how selected statistics vary with
the system parameter, without fitting separate
models at each condition.
Across Lorenz-63, KS, and Kolmogorov flow,
the Graph QCBM achieves the prescribed accuracy
threshold at unseen conditions after training on
only five parameter values.
Its performance relative to other conditional
architectures further supports the effectiveness
of graph evolution for representing these
parameter-dependent distributions.

The predictor experiments show that quantum priors can
improve autoregressive forecasts through regularization.
For Kolmogorov flow, regularization with a prior representing
the normalized two-dimensional vorticity power spectrum
also improves the mean vorticity field and the distribution
of vorticity values, extending its benefits beyond the
spectral statistics directly targeted by the penalty.

The Photonic QNN demonstrates a hardware
implementation of conditional prior learning
using a photonic quantum processor.
Experimentally measured quantum responses provide
nonlinear features, which are combined through
trainable classical readouts and interactions
to construct conditional probability distributions.
The resulting model generalizes to unseen parameter
values in both Lorenz-63 and Kolmogorov flow,
demonstrating the feasibility of integrating
photonic quantum hardware into conditional
prior learning.

Future work should investigate how the choice and
dimensionality of the statistical target affect predictor
accuracy.
Joint distributions over $D>2$ dimensions
could constrain dependencies omitted by the current
two dimensional targets.
At fixed resolution per dimension, however, the joint
grid grows exponentially with $D$.
A systematic comparison across target dimensions would
quantify the predictive benefit of these additional
dependencies alongside the trajectory data required to
estimate the targets and the model resources required to
represent them.
Relevant measures include forecast error, agreement with
reference statistics, trainable weight counts, circuit
depth, and measurement requirements.

A second direction is to examine how sampling affects
the contribution of prior regularization to predictor
optimization.
The supervised minibatch size affects the estimation of
the data-loss gradient, while the number of predictor
rollouts affects the estimation of the JS penalty and its
gradient.
Varying these sample counts independently, with a fixed
rollout horizon and consistent loss normalization, would
allow their effects on gradient variability and forecast
accuracy to be quantified.
This would test whether the prior is particularly
beneficial when data-loss gradients are noisy, and how
finite-sample error in the statistical penalty limits
that benefit.

\section{Conclusion}\label{sec:conclusion}
We develop a conditional quantum prior that learns how
selected invariant statistics depend on a system parameter.
One shared set of trainable weights provides statistical
references at unseen parameter values within the training
range without retraining.
For Lorenz-63, a Graph QCBM with 2,976 trainable weights
represents the selected statistics at five training and
96 unseen parameter values.
Sharing one prior across these 101 values reduces the
number of trainable weights by 84.7\% compared with fitting
a separate QIML QNN at each value.

With five training values per system, the Graph QCBM keeps
JS below $0.005$ at every tested unseen parameter value
in every run of the three coverage studies.
In the Lorenz-63 comparison at similar trainable weight
counts, its mean JS across 96 unseen values is 56.9\%
lower than that of the conditional MLP.

Regularization with a fixed Graph QCBM prior lowers test
trajectory MSE in all three evaluated systems.
For Lorenz-63 and KS, these improvements are achieved at
parameter values not used to train the prior.
For Kolmogorov flow at $Re=1000$, regularization reduces
MSE by 66.0\% over 318 forecast steps relative to the
predictor trained without the prior.
The mean vorticity field, distribution of vorticity values,
and power spectrum also agree more closely with the
reference data.

Finally, we demonstrate conditional prior learning
using quantum responses measured on photonic hardware,
combined with trainable classical processing.
The resulting model generalizes to unseen parameter
values in both Lorenz-63 and Kolmogorov flow.

\section{Acknowledgements}
\label{sec:ack}

This research is supported by the National Key R\&D Program of
China (Grant No.~2024YFB4504005 and 2024YFA1409300);
National Natural Science Foundation of China (NSFC)
(Grant Nos.~62235012, 12304342, 12574549, 12574542, and 125B1033);
Innovation Program for Quantum Science and Technology
(Grant Nos.~2021ZD0301500 and 2021ZD0300700);
Science and Technology Commission of Shanghai Municipality
(STCSM) (Grant Nos.~2019SHZDZX01, 24ZR1438700, 24ZR1430700,
and 24LZ1401500);
Shanghai Municipal Commission of Economy and Informatization (no. 2025-GZL-RGZN-02047);
and Frontier Technologies R\&D Program of Jiangsu
(Grant No.~SBF20250000094).

\bibliographystyle{unsrtnat}
\bibliography{qprior_references}
\appendix

\section{Graph Size and Distribution Accuracy}\label{app:architecture-results}\label{app:graph-protocol}
We compare four Graph QCBM configurations to assess
the tradeoff between model size and distribution accuracy.
A configuration with $L$ graph layers and $e$ edges
per layer is denoted by H$L$/$e$.
Each layer also contains sixteen coefficients for
encoding the system parameter on eight qubits,
giving $L(e+16)$ trainable weights.
Table~\ref{tab:compression} summarizes the results
on the Lorenz-63 benchmark.

\begin{table}[h!]
\centering\small
\begin{tabular}{lrr}
\toprule
Graph QCBM & Trainable weights
& Unseen test JS $(\times10^{-3})$ $\downarrow$ \\
\midrule
H8/480 & 3,968 & $0.550\pm0.036$ \\
H6/480 & 2,976 & $0.528\pm0.034$ \\
H6/360 & 2,256 & $0.621\pm0.052$ \\
H4/480 & 1,984 & $0.605\pm0.076$ \\
\bottomrule
\end{tabular}
\caption{Graph QCBM size and unseen Lorenz-63
accuracy in ideal simulation.
JS is averaged over 96 unseen parameter values
for each seed; entries report the mean and sample
standard deviation across five seeds.}
\label{tab:compression}
\end{table}

Among the tested configurations, H6/480 achieves
the lowest observed mean unseen JS while using
25\% fewer trainable weights than H8/480.
Its mean JS is 4.0\% lower than that of H8/480.
We therefore adopt H6/480 for the main experiments.

Reducing the graph to 360 edges per layer decreases
the number of trainable weights by 24.2\% relative
to H6/480, with an approximately 18\% increase
in mean unseen JS.
Reducing the number of layers from six to four
decreases the weight count by 33.3\% and increases
mean unseen JS by 14.6\%.
These smaller configurations provide alternatives
when model size is a greater constraint.
Further compilation studies are needed to determine
how these reductions affect native gate counts
and circuit depth.

All configurations use the same five training
parameter values, four unseen validation values,
and 96 independent test values within
$\rho\in[26,38]$.

For the 360-edge configuration, we score each edge
by the median increase in training JS when that
edge is removed from the fitted H6/480 models.
We construct a maximum spanning tree using these
scores to preserve connectivity between all
statistical bins.
The remaining edges are then added in descending
score order until the graph contains 360 edges.
All five runs use the resulting graph topology.

\section{Differentiating the Statistical Regularizer}\label{app:predictor-protocol}
Predictor regularization requires gradients through
statistics estimated from the predictor rollout.
For Lorenz-63 and KS, hard bin assignments are piecewise
constant in the observable values and therefore have
zero gradients almost everywhere.
We use triangular bin weights as a differentiable
approximation~\citep[Eq.~(2)]{ustinova2016}.

At optimization step $m$, let
$\mathbf z_n(\omega_m)\in\R^D$, $n=1,\ldots,N$,
denote observable samples obtained from the predictor
rollout.
For the prior's grid of $M$ bins with centers
$\mathbf c_i$, we use a triangular kernel with
half-width $\Delta_d$ along coordinate $d$:
\begin{align}
 C_i(\omega_m)
 &=
 \sum_{n=1}^{N}
 \prod_{d=1}^{D}
 \left[
 1-\frac{|z_{nd}(\omega_m)-c_{id}|}{\Delta_d}
 \right]_+,
 \nonumber\\
 \widehat p_{\omega_m}(i)
 &=
 \frac{C_i(\omega_m)+\varepsilon}
 {\sum_{j=1}^{M}C_j(\omega_m)+M\varepsilon},
 \qquad \varepsilon=10^{-8}.
 \label{eq:soft-histogram}
\end{align}
Here $[a]_+=\max(a,0)$ and $D=2$ for both joint
histogram targets.
The pseudocount $\varepsilon$ ensures a positive
normalization denominator.
Equation~\eqref{eq:soft-histogram} supports
backpropagation of the JS penalty through the estimated
histogram to the predictor weights $\omega_m$.
The learned prior remains fixed.

For Kolmogorov flow, the statistical table is constructed
from the normalized Fourier power of the predicted
vorticity fields.
This construction supports backpropagation without
histogram smoothing.

\section{Experimental Protocols}\label{app:protocol}

\subsection{Data Generation and Prior Training}
\label{sec:protocol}
\label{app:coverage-results}
\label{app:coverage-grids}

We generate the trajectory data used for prior training
and evaluation by numerical integration of the three
dynamical systems.
Empirical targets are constructed using the observables
and normalization conventions in Section~\ref{sec:targets}.
Within each experiment, all models use the same empirical
targets and statistical grid.

\paragraph{Lorenz-63.}
We integrate the Lorenz equations using the fourth-order
Runge--Kutta method with time step $0.01$ for 70,000 steps,
discarding the first 10,000 steps.
For the conditional architecture comparison and coverage
study, each empirical target is estimated from eight
trajectories at the corresponding parameter value.
Training and evaluation use independent trajectory ensembles.

\paragraph{Kuramoto--Sivashinsky.}
We integrate KS on 512 spatial points using a Fourier
pseudospectral method with two-thirds dealiasing and
fourth-order exponential time differencing
Runge--Kutta integration~\citep{kassam2005}.
The time step is $0.05$.
After discarding the transient up to $t=6000$, we sample
the solution at unit intervals through $t=10000$.
Each training parameter value has 16 training and
16 validation trajectories.
Each seen or unseen test parameter value has
32 test trajectories.

\paragraph{Kolmogorov flow.}
We generate trajectories using the JAX-CFD Fourier
pseudospectral solver~\citep{dresdner2022}, with
Crank--Nicolson integration for the linear terms and
Runge--Kutta integration for advection.
Simulations use a $1024\times1024$ grid on $[0,2\pi]^2$,
forcing amplitude $A=1$, forcing wavenumber $k_f=8$,
and linear drag coefficient $0.1$, as in
Eq.~\eqref{eq:kf}.
We project the velocity fields to a $64\times64$ grid
in Fourier space and store 2,400 frames at intervals
of $0.1$.
The spectral targets use the final 1,600 frames of
each trajectory.
Each training parameter value has eight training and
eight validation trajectories.
Each unseen evaluation value has eight trajectories.

\paragraph{Training coverage.}

Using these datasets, we examine how the number
of training parameter values affects the accuracy
of the Graph QCBM.
For Lorenz-63 and KS, we consider 2, 3, 5, 9, and 17
training conditions, while for Kolmogorov flow we
consider 2, 3, and 5.
Within each system, the model architecture, statistical
grid, and unseen evaluation values remain fixed.
Each training design is repeated with three model seeds.

\begin{figure}[h!]
\centering
\includegraphics[
width=\linewidth,
height=0.82\textheight,
keepaspectratio
]{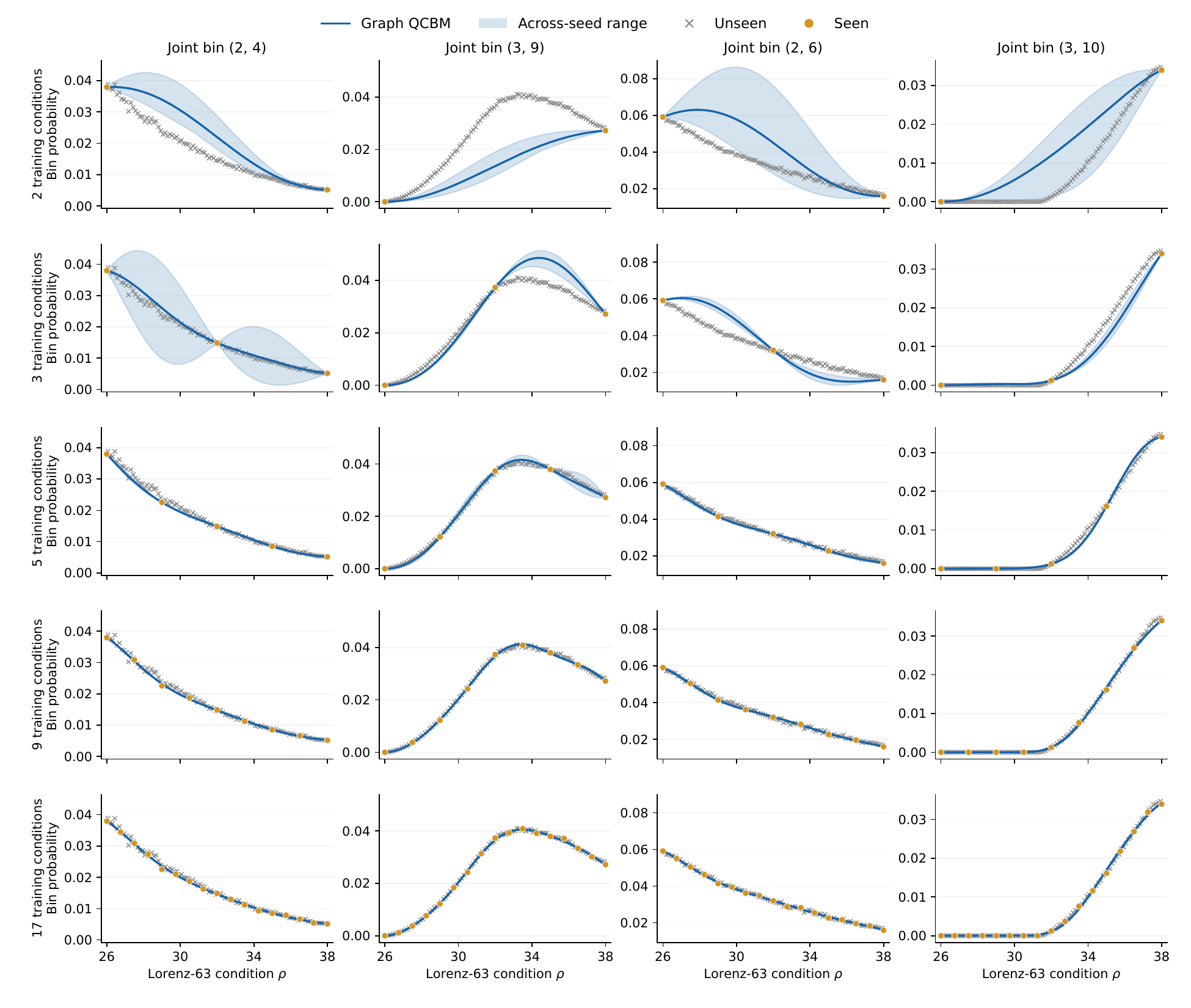}

\caption{Effect of training coverage on Lorenz
bin probabilities.
Rows correspond to 2, 3, 5, 9, and 17 training
conditions; columns show four selected bins.
Blue curves and shading indicate the mean and
range across three seeds.
Orange circles and gray crosses denote training
and unseen empirical targets, respectively.}
\label{fig:condition-responses}
\end{figure}

Figure~\ref{fig:condition-responses} illustrates the
effect of training coverage on four selected Lorenz
bin probabilities.
With fewer training conditions, the predicted curves
show substantial deviations from the empirical targets.
As the number of training conditions increases, the
predictions more closely follow the changing bin
probabilities, including at unseen parameter values.

Table~\ref{tab:coverage-full} reports the maximum
unseen JS taken over all
unseen evaluation conditions and three model seeds.
This provides a direct assessment of the worst-case
accuracy across the evaluated conditions and runs.

\begin{table}[h!]
\centering\small
\begin{tabular}{lrr}
\toprule
System & $N_{\rm train}$ & Maximum unseen JS $\downarrow$ \\
 & & $(\times10^{-3})$ \\
\midrule
Lorenz-63 & 2 & 121.481 \\
 & 3 & 23.426 \\
 & 5 & 2.804 \\
 & 9 & 0.863 \\
 & 17 & 0.629 \\
\midrule
KS & 2 & 102.184 \\
 & 3 & 22.316 \\
 & 5 & 1.283 \\
 & 9 & 0.0196 \\
 & 17 & 0.0094 \\
\midrule
Kolmogorov flow & 2 & 5.222 \\
 & 3 & 0.316 \\
 & 5 & 0.117 \\
\bottomrule
\end{tabular}

\caption{Maximum unseen JS in simulation for
different numbers of training conditions.}
\label{tab:coverage-full}
\end{table}

The maximum unseen JS decreases as training coverage
increases in all three systems.
With five training conditions, it falls below the
prescribed threshold of $0.005$ for every evaluated
condition and training run.
Increasing coverage to 9 or 17 conditions further
improves the worst-case accuracy for Lorenz-63 and KS.
These results support the use of five training
conditions in the main experiments while showing
that broader training coverage can improve accuracy.
\paragraph{Conditional model comparison.}
All four models in the Lorenz comparison share the
Graph QCBM optimization settings in
Section~\ref{sec:architecture-results} and select
checkpoints by minimum mean validation JS.
The three quantum models share the product reference
state and condition encoding in Eq.~\eqref{eq:encoding}.
The conditional QNN uses six blocks of twenty layers
of qubit rotations and CNOT gates on eight qubits,
following the QIML gate pattern~\citep{wang2026}.
The coin model uses six blocks of 32 steps, each
applying a trainable $\mathrm{SU}(4)$ transformation
to two auxiliary coin qubits followed by a shift
conditioned on their state~\citep{chang2026quantumwalks}.
Shifts reflect at the grid boundaries, and summing
probabilities over the coin states gives the
distribution over statistical bins.
The MLP takes normalized $\rho$ and the two bin
coordinates as inputs, uses widths
$3\to52\to51\to1$ with tanh hidden activations,
and normalizes its bin logits with a softmax.

\paragraph{QIML reference at a fixed parameter value.}
\label{app:qnn-control}
The separate QIML QNN is trained only on the empirical
target at $\rho=28$.
It retains the source circuit with eight qubits,
eight layers, 192 trainable weights, and initial
state $\ket{0}^{\otimes8}$~\citep{wang2026}.
The QNN shares the $p(r,z)$ histogram grid and
independent test target with the five H6/480
checkpoints in Appendix~\ref{app:architecture-results}.

\subsection{Predictor Training and Evaluation}\label{app:predictor-setup}
The Lorenz-63 predictor experiments use trajectory
data generated in this work.
For KS and Kolmogorov flow, we use the datasets
provided with QIML~\citep{wang2026}.
For each system, the reported comparison uses
one trained predictor with the selected prior
regularization strength and one trained predictor
without the prior.
\paragraph{Lorenz-63.}
At $\rho=28$, we follow the predictor architecture
and distribution matching objective of
Botvinick-Greenhouse et~al.~\citep{botvinick2025delay}.
The data objective uses energy distance to compare
predicted and observed distributions of states and
of delayed samples of the $x$ coordinate.
Our additional JS penalty compares the joint
distribution of $(r,z)$ with the learned prior,
using 50 predicted states per initial condition.

We compare regularization strengths
$\alpha_{\rm reg}\in\{0.01,0.1,1\}$ after
10,000 optimization updates.
Validation MSE over 110 forecast steps selects
$\alpha_{\rm reg}=0.01$.
Trajectory MSE is reported over 500 forecast steps
from 128 test initial states.
Statistical evaluation uses 10,000-step rollouts
from 64 test initial states, with distributional
JS computed over steps 5,001--10,000.

\paragraph{Kolmogorov flow.}
At $Re=1000$, we use the predictor architecture
and data objective from QIML~\citep{wang2026}
for vorticity fields on a $64\times64$ grid.
The dataset contains 42 training, 20 validation,
and 20 test trajectories.
Both predictors train for 100 epochs with batch
size 32 and identical initial weights.
The JS penalty uses a normalized two-dimensional
vorticity power spectrum estimated from ten
predicted frames per trajectory.

We compare regularization strengths
$\alpha_{\rm reg}\in\{0.01,0.1,1\}$ using
validation MSE over 140 forecast steps at epoch 100.
This selects $\alpha_{\rm reg}=0.01$.
Test evaluation uses 318 forecast steps from
each of the 20 test trajectories.

The radial plots sum power over modes grouped by
wavenumber magnitude.
For mean-field MSE, we first average the predicted
and reference vorticity fields separately over
time and trajectories, then compute the spatial
MSE between the resulting mean fields.
Vorticity distributions use 128 bins spanning
reference percentiles 0.1--99.9 with 5\% margins,
plus two overflow bins.

\paragraph{Kuramoto--Sivashinsky.}
At $\nu=1$, we use the predictor architecture
and data objective from QIML~\citep{wang2026}
on a grid of 512 spatial points.
A random trajectory split assigns 960 trajectories
to training, 120 to validation, and 120 to testing.

We set $\alpha_{\rm reg}=0.01$ and estimate the
joint derivative distribution for the JS penalty
from ten predicted frames per trajectory.
Both predictors train for 100 epochs with batch
size 400.
Evaluation uses their final checkpoints on the
same 120 test trajectories.
Trajectory MSE is reported over 500 forecast steps.

\subsection{Photonic Response Data and Model Configuration}
\label{app:photonic-protocol}

We use detection probabilities measured on photonic
quantum hardware across seven distinct optical
circuit configurations.
For each configuration, measurements are recorded
at 100 uniformly spaced optical phases over
$[0,2\pi)$, with 44--63 retained detection outcomes.
These measurements provide 373 response curves in total.

Each configuration consists of squeezed-vacuum
light passing through an interferometer with
fixed optical transformations and a programmable
phase.
Varying this phase changes the interference
and the probabilities of the detection outcomes,
while the remaining optical settings are fixed.
Each retained detection pattern therefore defines
a probability response as a function of phase.
The detection probabilities are normalized
over the retained outcomes at each optical phase.
Periodic linear interpolation provides responses
at intermediate phases.
The interpolated hardware responses remain fixed
throughout training and evaluation.

For both systems, $d_{\rm axis}=12$.
We use $d_\lambda=8$ for Lorenz-63 and
$d_\lambda=4$ for Kolmogorov flow,
resulting in 1,862 and 1,234 trainable
weights, respectively.

The underlying optical circuits follow the
Gaussian boson sampling (GBS) paradigm.
For an ideal pure Gaussian state with zero
displacement and photon-number-resolving detection,
the probability of an output photon-number
pattern $\mathbf n=(n_1,\ldots,n_{N_{\rm modes}})$
is given by
\begin{equation}
\pi_{\mathbf n}(\varphi)
=
\frac{
\left|\operatorname{Haf}\!\left[
B_{\mathbf n}(\varphi)\right]\right|^2
}{
\sqrt{\det Q(\varphi)}
\prod_{j=1}^{N_{\rm modes}} n_j!
}.
\end{equation}
Here $N_{\rm modes}$ is the number of optical modes,
$B_{\mathbf n}(\varphi)$ is obtained by repeating
the rows and columns of the symmetric matrix
$B(\varphi)$ according to the photon occupation
numbers in $\mathbf n$, and $Q(\varphi)$ is the
covariance matrix of the Gaussian state's
Husimi $Q$ function.

\end{document}